\documentclass[%
reprint,superscriptaddress,amsmath,amssymb,twocolumn,nofootinbib,aps,pra
 ]{revtex4-2}
\usepackage[columnwise,mathlines]{lineno}
\usepackage{graphicx}% Include figure files
\usepackage{appendix}
\usepackage{dcolumn}% Align table columns 
\usepackage{bm}
\usepackage{braket}
\usepackage{amsmath}
\usepackage{amssymb}
\usepackage{multibib}
\usepackage{graphicx}
\usepackage[colorlinks=true, allcolors=blue]{hyperref}
\usepackage{mathtools}
\usepackage{enumerate}
\usepackage{textgreek}
\begin{document}

\title{Collective light-matter interaction in plasmonic waveguide quantum electrodynamics}

\author{Zahra Jalali-Mola}
\affiliation{School of Physics and CRANN Institute, Trinity College Dublin, Dublin 2, Ireland}

\author{Saeid Asgarnezhad-Zorgabad}
\email{sasgarnezhad93@gmail.com}
\affiliation{School of Physics and CRANN Institute, Trinity College Dublin, Dublin 2, Ireland}

\date{\today}
\begin{abstract}
Rabi oscillations characterize light–matter hybridization in the waveguide quantum electrodynamics~(WQED) framework, with their associated decay rates reflecting excitation damping, yet their behavior remains unresolved when collective emitters are coupled to a collective waveguide mode. This scenario reveals a conceptually novel collective-light-collective-matter interaction, realizable when a timed-Dicke state of subwavelength emitters couples to a slow, delocalized surface-plasmon mode, forming a hybridized plasmon-polariton~(HPP). The HPP acquires its directionality from the timed-Dicke state via momentum matching. It also exhibits plasmonic characteristics, with excitation frequencies following the surface-plasmon dispersion relation. We obtain a Rabi oscillation and a long-time decay that describe the HPP and use them to reveal weak- and strong-coupling regimes through the emergence of normal-mode splitting. By performing a finite-time Lyapunov-exponent analysis, we show that the HPP also exhibits instantaneous decay and numerically identify three distinct decay regimes: early-time rapid, transient-time oscillatory, and long-time classical. Finally, by analyzing the emission spectrum, we observe an anticrossing of the peak doublets—a feature also seen in cavity QED setups—which originates from quantum vacuum effects and the resulting non-Markovian HPP evolution in our WQED.  
\end{abstract}

\maketitle

\section{Introduction}
Collective states play a central role in quantum electrodynamics~(QED) because they evolve as a single coherent mode, enabling robust transport by providing long-range coherence~\cite{RevModPhys.95.015002}. The interaction between collective-light and collective-matter states can further enhance light–matter coupling, making the resulting hybridized excitations pivotal for photon control in optical communication networks~\cite{Walther_2006,kimble2008quantum,RevModPhys.94.041003}. Surface plasmons and timed-Dicke states~(TDS)~\cite{PhysRevLett.102.143601} are established collective excitations of light and matter that emerge at metallic interfaces and in atomic ensembles, respectively, and provide natural platforms for realizing collective light–matter interactions. Previous studies have demonstrated strong coupling between surface plasmons and quantum emitters~\cite{fang2015nanoplasmonic,chikkaraddy2016single,Stockman_2018,meng2021optical,https://doi.org/10.1002/adom.202001520,PhysRevB.82.075427,PhysRevB.84.075419,PhysRevLett.110.126801,Delga_2014,doi:10.1021/acsphotonics.7b00475,JalaliMolaAsgarnezhadZorgabad+2021+3813+3821,PhysRevA.98.013825,PhysRevA.99.051802,PhysRevA.104.L051503,doi:10.1021/acsphotonics.9b00193,PhysRevB.98.195415,PhysRevB.100.205413,PhysRevResearch.1.023027,gonzalez2024light,Liu:25}; however, the interaction between a surface plasmon and a TDS, constitutes a genuinely collective-light–collective-matter process, is a novel concept that remains largely unexplored. In this work, we investigate this interaction and explore spatial-temporal features of the resultant hybridized state.

\begin{figure}
    \centering
    \includegraphics[width=0.9\linewidth]{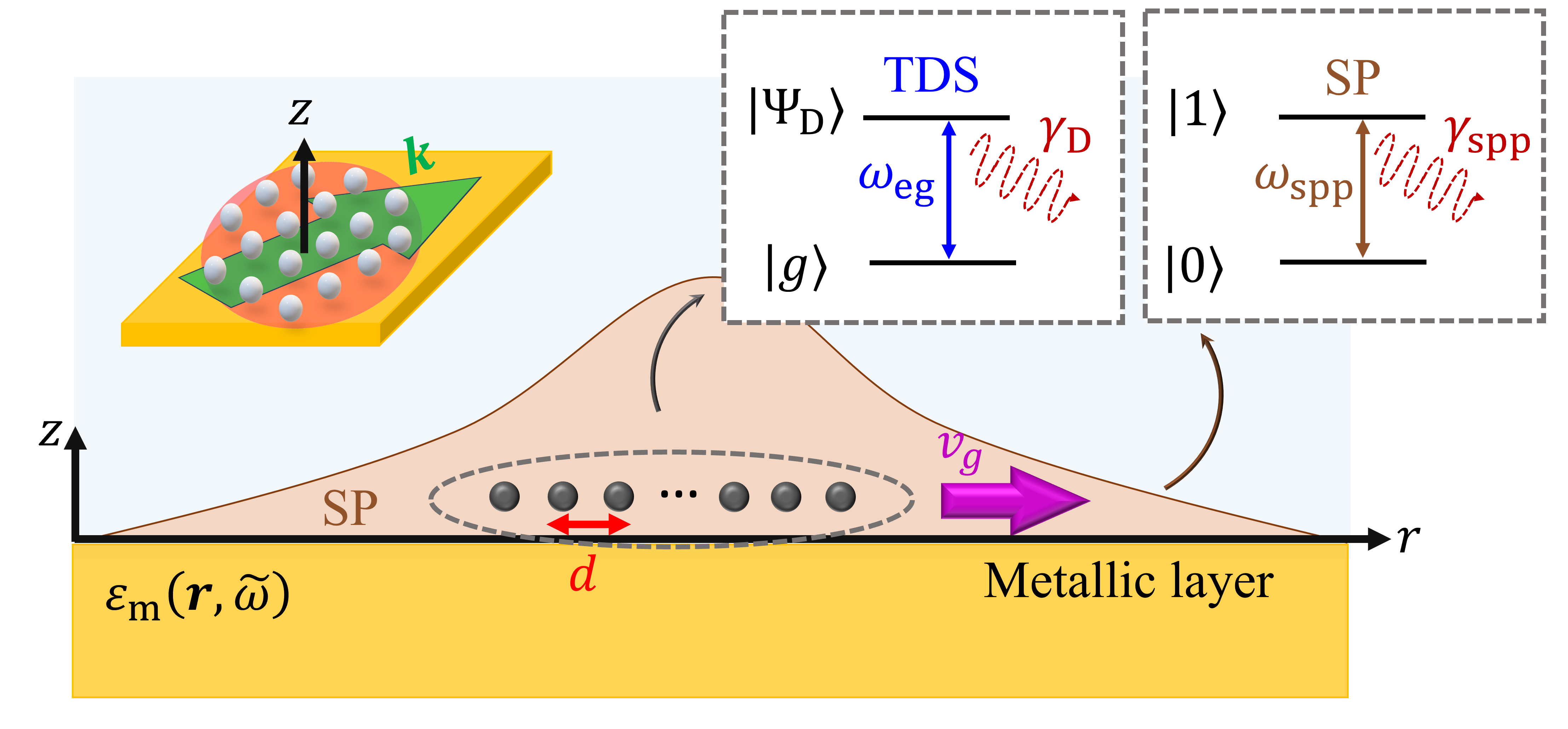}
    \caption{Interaction between TDS and surface-plasmon field: An ensemble of $N$ QEs, equidistantly spaced by $d$ and prepared in a TDS, have ground~($\ket{g}$) and excited~($\ket{\Psi_\text{D}}:=\ket{\Psi_\text{D}(N|k)}$) states, with frequency separation $\omega_\text{eg}$. The TDS is prepared by an external laser $\bm{k}$ and is situated on top of a metallic layer with optical properties $\varepsilon_\text{m}(\bm{r}',\tilde{\omega})$ and $\mu_\text{m}(\bm{r}',\tilde{\omega})=1$, where $\bm{r}'$ includes the lower half-plane $z<0$. Interaction between the TDS and the surface plasmon evolves as the HPP in the direction $\bm{k}$ at frequency $\omega_\text{eg}$. The surface plasmon with vacuum $\ket{0}$ and single-photon $\ket{1}$ states has a group velocity $v_\text{g}$, excitation frequency $\omega_\text{spp}:= \Re[\tilde{\omega}_\text{spp}]$, and is delocalized enough to cover the QE ensemble. The top-left inset shows the top view of this interaction, while the top-right inset shows the energy diagrams of the TDS and surface plasmon.}

    \label{fig:Geometry_Waveguide}
\end{figure}

Our analysis of this interaction demonstrates that the hybridized state~[here termed hybridized plasmon-polariton~(HPP)] exhibits collective behavior and can be uniquely characterized by a collective frequency, a long-time decay, and an instantaneous decay. Building on this, we justify that our collective-collective interaction shares similarities and distinctions with the interaction between a single-mode field and a single emitter in the QED framework. Similar to standard QED setups, we identify weak, weak-to-strong, and strong coupling regimes that emerge from the competition between collective frequency and long-time decay. 

Nevertheless, instantaneous decay, originating from the non-Markovian evolution of the HPP, is the key distinction between our proposed interaction and conventional QED frameworks. Specifically, we show that this decay features three distinct regimes: an early-time fast quantum-like, a transient-time non-classical oscillatory, and a long-time classical decay. Establishing similarities to QED setups and demonstrating various coupling strengths is our technical novelty, whereas employing the finite-time Lyapunov exponent to uncover the existence of instantaneous decay reflects our methodological novelty. The rest of this paper is organized as follows. In \S~\ref{Sec:Model_Math_Methods}, we present the model and the mathematical methods used to describe the hybrid plasmonic WQED. In \S~\ref{Sec:Results_Discussion}, we discuss the temporal and spectral evolution of the HPP state. Finally, in \S~\ref{Sec:Conclusion_Outlook},
we summarize our findings and discuss possible future directions.    

\section{\label{Sec:Model_Math_Methods} Model, mathematics and methods}
In this section, we introduce the model, mathematics, and methods, beginning with the collective light–matter interaction as a conceptual framework and its physical realization in terms of the interaction between surface plasmons and a TDS in \S~\ref{Sec:Concept_Framework}. We present a qualitative description of our plasmonic WQED system in \S~\ref{Sec:Model}, and describe the mathematical framework of the HPP dynamics, namely the quantitative analysis of the interaction between a quantum plasmon mode and the TDS using a Schr\"odinger picture, in \S~\ref{Sec:Mathemtics}. We then elucidate our approach for solving the resulting dynamics in the complex-frequency plane and real wavenumber, and finally, test the feasibility of our framework in \S~\ref{Sec:Methods_Feasibility}.

\subsection{\label{Sec:Concept_Framework} Conceptual framework}
The surface-plasmon field and the TDS can be described via a complex frequency $\tilde{\omega}_\text{spp}=\omega_\text{spp}+\text{i}\gamma_\text{spp}$ and $\tilde{\omega}_\text{eg}=\omega_\text{eg}+\text{i}\gamma_\text{D}$, where $\omega_\text{spp}$~($\omega_\text{eg}$) and $\gamma_\text{spp}$~($\gamma_\text{D}$) describe collective oscillation and long-time decay, respectively. The coupling between these two in the bare state can be understood as the interaction between a single emitter and a continuum of frequencies obeying the surface-plasmon dispersion relation. Discretizing this continuum leads to a large multimode emitter–field problem, which is computationally intractable and makes a bare-state analysis difficult. Despite this complexity, we can anticipate non-Markovian behavior of the interacting system. 

In this framework, the dynamics can be equivalently described by either an integrodifferential equation or a damped-harmonic oscillator equation. Specifically, in the resonant case~($\omega_\text{spp}=\omega_\text{eg}$) and for a slowly propagating surface-plasmon mode, the interacting system is expected to exhibit various coupling regimes if a well-defined interaction parameter~(in this case, the effective collective vacuum Rabi frequency, which we term \textit{effective frequency} $\Omega_\text{s}$) becomes comparable to the maximum long-time decay rate of the system, $\max\{\gamma_\text{spp},\gamma_\text{D}\}$. We anticipate pure-decay dynamics~(weak coupling) for $\Omega_\text{s}<\max\{\gamma_\text{spp},\gamma_\text{D}\}$ while oscillatory dynamics~(strong coupling) should appear for $\Omega_\text{s}>\max\{\gamma_\text{spp},\gamma_\text{D}\}$. These insights constitute the conceptual framework of this study. In what follows, we justify the validity of this framework for $\gamma_\text{D}=0$ even for highly non-Markovian HPP dynamics. 

\subsection{\label{Sec:Model} Model}
To model the interaction between the surface-plasmon field and the TDS, we consider a metallic layer\footnote{Here, we consider a gold layer;
however, we predict that our framework remains valid for metallic two-dimensional material such as graphene~\cite{JALALIMOLA2019220}, double-layer schemes~\cite{PhysRevB.98.235430} supporting surface-plasmon field.} located at $z<0$, and an ensemble of $N$ quantum emitters~(QEs) placed in its vicinity~\cite{doi:10.1126/science.aah3752,doi:10.1126/science.aah3778,PhysRevX.8.041054}, forming a plasmonic WQED, as illustrated in Fig.~\ref{fig:Geometry_Waveguide}. We assume the metallic layer is non-magnetic~($\mu_\text{m}(\bm{r}',\omega)=1$), whose electric permittivity $\varepsilon_\text{m}(\bm{r}',\tilde{\omega})$ is described by the Drude model~(see Eq.~\eqref{Eq:Drude} of the appendix \ref{Sec:General_SPP} and also Ref.~\cite{PhysRevB.95.115444}). We further assume that the QEs are identical two-level systems, each with excited~($\ket{e}$) and ground states~ ($\ket{g}$), transition frequency $\omega_\text{eg}$, and a dipole moment $\bm{p}$, which is located at position $\bm{r}_i$. To realize the TDS, we assume that the QEs are placed at equal spacing $d$, at the same height $z$ above the interaction interface, and arranged in a two-dimensional subwavelength lattice satisfying $Nd\ll\lambda_\text{eg}$~\cite{masson2022universality,PhysRevResearch.4.023207}. 

In this apparatus
(see the two-dimensional $z$-$r$ cut in Fig.~\ref{fig:Geometry_Waveguide}), the metallic layer supports a surface-plasmon field with excitation frequency $\tilde{\omega}_\text{spp}$ and group velocity $v_\text{g}$, whereas QEs evolve as the TDS. Hence, collective-light-collective-matter interaction can be realized in our proposed WQED setup. We assume that the QEs are already prepared in the TDS
\begin{equation}
\ket{\Psi_\text{D}(N|k)}=(1/\sqrt{N})\sum_{i=1}^N e^{\text{i}\bm{k}\cdot\bm{r}_i}\ket{e_i}\underset{i\neq j}\otimes\ket{g_j,\{0\}},   
\label{Eq:Timed_Dicke_State}
\end{equation}
where $\bm{k}$ is the wavenumber of the external field and $\ket{g_j,\{0\}}:=\ket{g_j}\otimes\ket{0}$~(with $\ket{0}$ denoting the vacuum-state of the surface-plasmon field; see appendix \S~\ref{Approach:Schrodinger})\footnote{We note that this external field does not contribute to the interaction; however, its wavenumber determines the TDS direction}. Here, we do not investigate the details of the TDS preparation, as it requires accounting for many-body interactions, their coupling to the plasmon field during the preparation process, surface-plasmon inhomogeneities, and the properties of the illumination field. Developing such a many-body framework is beyond the scope of this work. Nevertheless, the hybridization between the surface plasmon and the TDS evolves as HPP, whose emission spectrum can then be measured with a detection system of sensitivity $\omega_\text{p}\simeq0.01$~eV.

\subsection{\label{Sec:Mathemtics} Mathematical description of the hybrid plasmonic waveguide}
We now use these insights to provide a quantitative description of our waveguide.
To this aim, we construct the total Hamiltonian of the system $H$, which comprises the QEs' Hamiltonian $H_\text{e}$, the field Hamiltonian $H_\text{f}$, and the interaction Hamiltonian $H_\text{int}$. To describe QEs, we introduce the $i$-th emitter raising~(lowering) operators as $\sigma_{+}^{(i)}:=\ket{e_i}\bra{g_i}$~($\sigma_{-}^{(i)}=\sigma_{+}^{(i)\dagger}$). Furthermore, we assume the surface plasmon as a bosonic field, and describe it using creation~(annihilation) operators $\bm{f}^{\dagger}(\bm{r},\tilde{\omega})$~($\bm{f}(\bm{r},\tilde{\omega})$), whose components satisfy the commutation relation
\begin{equation*}
    \left[f_\imath(\bm{r},\tilde{\omega}),f_\jmath^{\dagger}(\bm{r}',\tilde{\omega}')\right]=\delta_{\imath\jmath}\delta(\bm{r}-\bm{r}')\delta(\tilde{\omega}-\tilde{\omega}'),
\end{equation*}
for $\imath,\jmath\in\{x,y,z\}$~\cite{scheel2008macroscopic}, where $\tilde{\omega}$ is the complex probe frequency. Within the WQED framework and under the dipole approximation, the interaction Hamiltonian can be expressed in terms of field’s and QEs’ operators as $$H_\text{int}=-\sum_{i=1}^{N}(\sigma_{+}^{(i)}+\sigma_{-}^{(i)})\bm{p}_i\cdot\bm{E}(\bm{r}_i),$$ where $\bm{E}(\bm{r}_i)$ denotes the quantized electric field at the position of the $i$th QE. 

Here, $\bm{E}(\bm{r}_i)$ can be written as $\bm{E}(\bm{r}_i)=\int d\tilde{\omega}\bm{E}(\bm{r}_i,\tilde{\omega})$, with $\bm{E}(\bm{r}_i,\tilde{\omega})$ expressed in terms of the quantum current noise $\bm{j}_\text{n}(\bm{r}',\tilde{\omega})$ as~\cite{scheel2008macroscopic}
\begin{equation}
    \bm{E}(\bm{r}_i,\tilde{\omega})=-\text{i}\mu_0\tilde{\omega}\int d^3\bm{r}'~\bm{G}(\bm{r}_i,\bm{r}',\tilde{\omega})\cdot\bm{j}_\text{n}(\bm{r'},\tilde{\omega}),
    \label{Eq:Quantum_Pleasemonic_Field}
\end{equation}
where\footnote{In this work, we define $\Im[\cdot]$~($\Re[\cdot]$) as imaginary~(real) part of a complex number, respectively.}
\begin{equation*}
   \bm{j}_\text{n}(\bm{r}',\tilde{\omega})=\tilde{\omega}\sqrt{\hbar\varepsilon_0\varepsilon_\text{i}/\pi}\bm{f}(\bm{r}',\tilde{\omega}) 
\end{equation*}
with $\varepsilon_\text{i}:=\Im[\varepsilon_\text{m}(\bm{r}',\tilde{\omega})]$~\cite{Philbin_2010}, and $\bm{G}(\bm{r},\bm{r}';\tilde{\omega})$ is the system’s total Green’s function. We note that $\bm{G}(\bm{r},\bm{r}';\tilde{\omega})$ can be decomposed into the surface-plasmon Green’s function $\bm{G}_\text{sp}(\bm{r},\bm{r}';\tilde{\omega})$ and the regularized Green’s function $\bm{G}_\text{reg}(\bm{r},\bm{r}';\tilde{\omega})$. The former represents the pole contribution, giving rise to collective matter (surface-plasmon) excitation, whereas the latter accounts for other near-field effects such as creeping waves and a continuum of evanescent fields~\cite{PhysRevB.79.195414}. Since the aim of the present work is to explore collective light–collective matter interactions, we retain only $\bm{G}_\text{sp}(\bm{r},\bm{r}';\tilde{\omega}) := \bm{G}(\bm{r},\bm{r}';\tilde{\omega})$ and defer the inclusion of $\bm{G}_\text{reg}(\bm{r},\bm{r}';\tilde{\omega})$ to future work.

We then employ aforementioned definitions to express $H$ as
\begin{equation}
    H = H_\text{e}+H_\text{f}+H_\text{int}.
\end{equation}
To explore the collective excitations of the interacting system, we adopt the interaction picture with respect to $H_\text{e}+H_\text{f}$. In this rotated frame, we express QEs' and fields' operators as
\begin{align}
    \sigma_+^{(i)} &\mapsto \sigma_+^{(i)} \exp\{ i\omega_{\mathrm{eg}} t \},
    \label{Eq:Rotated_Sigma}
    \\
    \mathbf f(\mathbf r',\tilde{\omega}) &\mapsto \mathbf f(\mathbf r',\tilde{\omega}) \exp\{- i\tilde{\omega} t\},
    \label{Eq:Rotated_f}
\end{align}
respectively.
Inserting the field operator into $\bm{j}_\text{n}(\bm{r}',\tilde{\omega})$,
substituting the results into Eq.~\eqref{Eq:Quantum_Pleasemonic_Field}, and then plugging the resulting expression into $H_\text{int}$ we obtain 
\begin{align}
    H_\text{int}&=\text{i}\mu_0\sqrt\frac{\hbar\varepsilon_0}{\pi}\int_{i,\tilde{\omega},\bm{r}'}\sqrt{\varepsilon_\text{i}}\sigma_{+}^{(i)}\bm{p}_i\cdot\bm{G}(\bm{r}_i,\bm{r}',\tilde{\omega})\cdot\bm{f}(\bm{r}',\tilde{\omega})\nonumber\\&\times\exp\{-\text{i}(\tilde\omega-\omega_\text{eg})t\}+\text{c.c.},
    \label{Eq:Hamiltonian_Int}
\end{align}
where we use the following short-hand notation $$\sum_{i=1}^{N}\int d\tilde{\omega}\tilde{\omega}^2\int d^3\bm{r}':=\int_{i,\tilde{\omega},\bm{r}'}.$$ 

The Hilbert space of the interacting system can be decomposed into a state where all QEs are in their ground states and the surface plasmon is in the vacuum state
\begin{equation}
    \ket{\psi_\text{G}}=\bm{f}^{\dagger}(\bm{r}',\tilde{\omega})\ket{\mathcal{G},\{0\}},
\end{equation}
for $\ket{\mathcal{G},\{0\}}:=\ket{g_1,g_2,\ldots,g_N}\otimes\ket{0}$ denotes the ground state of the system, a delocalized TDS~(the maximally symmetric collective state $\ket{\Psi_\text{D}(N|\bm{k})}$ as given by Eq.~\eqref{Eq:Timed_Dicke_State}), and other collective states orthogonal to the TDS, which we denote as~($\ket{\Psi_{l\perp}}$). As TDS~[Eq.~\eqref{Eq:Timed_Dicke_State}] is a single-excitation collective state of QEs, our treatment is valid only in the single-excitation regime. We leverage the aforementioned Hilbert space to express the system wavefunction as  
\begin{align}
  \ket{\Psi(t)}=&\alpha_\text{D}(t)\ket{\Psi_\text{D}(N|k)}+\sum_{l=1}^{N-1}\alpha_{l\perp}(t)\ket{\Psi_{l\perp}}\nonumber\\&+\int d\tilde{\omega}\int d^{3}\bm{r}'\xi(\bm{r}',t,\tilde{\omega})\ket{\psi_\text{G}}.
  \label{Eq:Wavefunction}
\end{align}
In this equation, $\alpha_\text{D}(t)$ represents the transition amplitude of the $\ket{\Psi_\text{D}(N|k)}$ state, whose dynamics are directly related to the HPP, characterizing the hybridization between the TDS and the surface plasmon. We obtain the dynamics of the TDS $\dot{\alpha}_\text{D}(t)$ by inserting Eq.~\eqref{Eq:Hamiltonian_Int} and Eq.~\eqref{Eq:Wavefunction} into the Schr\"odinger equation.

Our derivation of $\dot{\alpha}_\text{D}(t)$ is based on two key assumptions. First, we assume that the waveguide size along the $r$ direction exceeds the TDS’s excitation wavelength, $\lambda_\text{eg}$, by a few times; therefore, the interaction surface exhibits in-plane translational symmetry. We leverage this property and the multiplication identity $$(\tilde{\omega}/\text{c})^2\int d^3\bm{r}' \varepsilon_\text{i}\bm{G}(\bm{r}_i,\bm{r}';\tilde\omega)\cdot\bm{G}^*(\bm{r}',\bm{r}_j;\tilde\omega)$$ to obtain a Fourier-space representation of the Green’s function $\Im[\bm{G}(\bm{q}_\parallel;\tilde{\omega})]$, where $\bm{q}_\parallel$ denotes in-plane symmetry~(see \S~\ref{Approach:Schrodinger} of the appendix and Ref.~\cite{PhysRevA.75.053813}). The second assumption concerns the finite number of QEs~\cite{PhysRevResearch.4.023207,masson2022universality} and the momentum-matching condition. It allows us to represent
\begin{equation*}
    \zeta(\bm{q}_\parallel,\bm{k}):=\sum_{ij=1}^N\exp\{\text{i}(\bm{q}_\parallel-\bm k)\cdot(\bm{r}_i-\bm{r}_j)\},
\end{equation*}
both as sum over QEs' spatial distribution and as a Gaussian distribution function centered at $\bm{k}$ with width $\mathcal{L}$, namely, 
\begin{equation}
    \zeta(\bm{q}_\parallel,\bm{k})\sim N^2\exp\{-\mathcal{L}^2(\bm{q}_\parallel-\bm{k})^2\}.
\end{equation}
In \S~\ref{Sec:General_SPP} of the appendix, we prove the convergence of these representations. In particular, for $\Omega_\text{s}=\gamma_\text{spp}$, we achieve convergence for a $16\times16$ QE lattice, independent of lattice structure. Using these assumptions and defining the emitter-emitter coupling as $$\mathcal{J}(\tilde{\omega},\bm{q}_\parallel)=\bm{p}_i\cdot\Im[\bm{G}(\bm{q}_\parallel;\tilde{\omega})]\cdot\bm{p}_j,$$ we derive $\dot{\alpha}_\text{D}(t)$ as
\begin{align}
    \dot{\alpha}_\text{D}(t)=&\int_0^t d\tau\int d\tilde{\omega}\int\frac{d^2\bm{q}_{\parallel}}{(2\pi)^2}
    \nonumber\mathcal{J}(\tilde{\omega},\bm{q}_\parallel)\\&\times\zeta(\bm{q}_\parallel,\bm{k})\alpha_\text{D}(\tau)
    \exp\left[\text{i}\left(\tilde{\omega}-\omega_\text{eg}\right)\left(\tau-t\right)\right],
    \label{Eq:Wavefunction_Momentum}
\end{align}
where $\dot{\alpha}_\text{D}(t):=\partial_t\alpha_\text{D}$ denotes the time derivative~(see appendix \S~\ref{Approach:Schrodinger} for the mathematical details of the derivation of Eq.~\eqref{Eq:Wavefunction_Momentum}.) 

\subsection{\label{Sec:Methods_Feasibility} Methods, Dynamics and Feasibility}
Building on the Fourier-optics formalism of the surface-plasmon field~\cite{PhysRevB.79.195414}, we express $\mathcal{J}(\tilde{\omega},\bm{q}_\parallel)$ in terms of a complex frequency and a real wavenumber. Assuming single-mode surface-plasmon excitation, $\mathcal{J}(\tilde{\omega},\bm{q}_\parallel)$ has a single pole at $$\tilde{\omega}_\text{spp}(\bm{q}_\parallel)=\omega_\text{spp}(\bm{q}_\parallel)+\text{i}\gamma_\text{spp}(\bm{q}_\parallel),$$ with residue $\bm{\mathcal{A}}(\bm{q}_\parallel)$~(see \S~\ref{Sec:General_SPP} for mathematical details). We further assume that the emitters' dipole moments are oriented along the $z$-axis; thereby, only the out-of-plane component of the Green tensor, $G_{z_{i}z_{j}}(\bm{q}_\parallel,\tilde{\omega})$, contributes to the interaction. Then $\Im[G_{z_{i}z_{j}}(\bm{q}_\parallel,\tilde{\omega})]$ for a single-pole plasmonic-field excitation has a Lorentzian lineshape with linewidth $\gamma_\text{spp}$ 
\begin{equation}
    \mathcal{J}(\bm{q}_\parallel,\tilde{\omega})=\frac{\gamma_\text{spp}}{2\pi}\frac{\mathcal{A}(\bm{q}_\parallel)}{(\tilde{\omega}-\omega_\text{spp}(q_\parallel))^2+\gamma_\text{spp}^2}.
    \label{Eq:Density_of_states}
\end{equation}

We substitute Eq.~\eqref{Eq:Density_of_states} into Eq.~\eqref{Eq:Wavefunction_Momentum} and perform integration over $\tilde{\omega}$ in the complex plane using the fact that the dominant contribution of $\mathcal{J}(\bm{q}_\parallel,\tilde{\omega})$ comes from $\tilde{\omega}=\tilde{\omega}_\text{spp}$. The subsequent integration over $\bm{q}_\parallel$ is then carried out by linearizing the dispersion relation as $$\tilde{\omega}(\bm{q}_\parallel)\approx\tilde{\omega}_\text{spp}+\bm{v}_\text{g}\cdot\bm{q}_\parallel,$$ and assuming $\bm{\mathcal{A}}(\bm{q}_\parallel)\approx\bm{\mathcal{A}}(\bm{k})$. We obtain $\dot{\alpha}_\text{D}(t)$ as
\begin{equation}
    \dot{\alpha}_\text{D}(t)=-\Omega_\text{s}\int_{0}^{t}d\tau \alpha_\text{D}(\tau)K(t-\tau), 
    \label{Eq:Dynamic}
\end{equation}
for
\begin{equation*}
K(t-\tau)= \exp\!\left[-\left(\frac{v_{\mathrm g}}{2\mathcal{L}}(t-\tau)\right)^2-\left(\gamma_{\mathrm{spp}}+\mathrm{i}\,\bm{k}\!\cdot\!\bm{v}_{\mathrm g}\right)(t-\tau)\right],
\end{equation*}
the memory kernel, and $\Omega_\text{s}\sim N\mathcal{J}(\tilde{\omega}_\text{spp},\bm{k})/(2\mathcal{L})^2$ the \textit{effective} frequency~(see \S~\ref{Approach:Schrodinger} for the mathematical details). Equation \eqref{Eq:Dynamic} describes the non-Markovian evolution of a timed Dicke state~(TDS) of a quantum-emitter array coupled to a surface-plasmon mode. In \S~\ref{Sec:Challenge_Dynamics} we have developed the master equation framework equivalent to Eq.~\eqref{Eq:Dynamic}. Our master equation, namely, Eq.~\eqref{Eq:Master_Eq} of the appendix, can be used to investigate HPP dynamics provided that the stability and convergence of the corresponding matrix-form integrodifferential equation are understood, which remains an open mathematical question~(see also appendix \ref{Sec:Master_Equation_USC_Regime} for remarks on the use of master equation).        

Now we test the feasibility of our theoretical framework by considering realistic parameters. We assume $\mathcal{L}=100~\mu\text{m}$, which is typically multiple times larger than the excitation wavelength $\lambda_\text{eg}$. We then use the Drude model~(see Eq.~\eqref{Eq:Drude} of the appendix \ref{Sec:General_SPP}) with background constant $\varepsilon_\infty=9$, plasma frequency $\hbar\omega_\text{pl}=9$~eV, damping rate $\hbar\gamma_\text{pl}=0.1$~eV, and set $\varepsilon_\text{g}=2.2$ as the background permittivity of the dielectric layer. Solving the dispersion relation of surface plasmon~\cite{PhysRev.182.539} for $\omega_\text{spp}=1.5$~eV, we achieve $|\bm{q}_\parallel|\approx0.012~\text{nm}^{-1}$,  $v_\text{g}=1.768\times10^{17}~\text{nm} \cdot\text{s}^{-1}$ and $\gamma_\text{spp}=5$~meV. Using these illustrative parameters and for $z_i=z_j=10$~nm, we obtain 
\begin{equation*}
    |\bm{G}_{z_iz_j}(\bm{q}_\parallel,\tilde{\omega})|\approx1\times10^{10}~\text{nm},
\end{equation*}
see \S~\ref{Sec:General_SPP} for more quantitative details. Here, we investigate the temporal–spectral evolution of $\dot{\alpha}_\text{D}(t)$ by neglecting TDS decay into free space and into the metal. Indeed, the TDS exchanges energy with the surface plasmon through the vacuum effect, and this field subsequently experiences Ohmic loss as described by the Drude model. Furthermore, using these values, the characteristic surface-plasmon timescale $\gamma_\text{spp}^{-1}$ becomes much shorter than the surface-plasmon stay-time, which is characterized by $\mathcal{L}/v_\text{g}$. Even in this simplified loss-free TDS case, we obtain Eq.~\eqref{Eq:Dynamic}, which contains a complex memory kernel. Considering additional loss channels for both the TDS and the plasmon would be more realistic, but it requires a master-equation approach, the analysis of which is beyond the scope of this work. 

\section{\label{Sec:Results_Discussion} Results and Discussion}
We present the results of our work in three sections. First, in \S~\ref{Sec:Result_Temporal_Spectral}, we analyze the temporal and spectral characteristics of the HPP obtained via numerical solution of Eq.~\eqref{Eq:Dynamic}. We then perform a Lyapunov analysis and introduce the different HPP decay regimes in \S~\ref{Sec:Lyapunov_Exponent}. Finally, we discuss the anticrossing of the lower and upper branches of the HPP in \S~\ref{Sec:Discussion}.

\subsection{\label{Sec:Result_Temporal_Spectral} Temporal and spectral analysis}
\begin{figure}
    \centering
    \includegraphics[width=1.\linewidth]{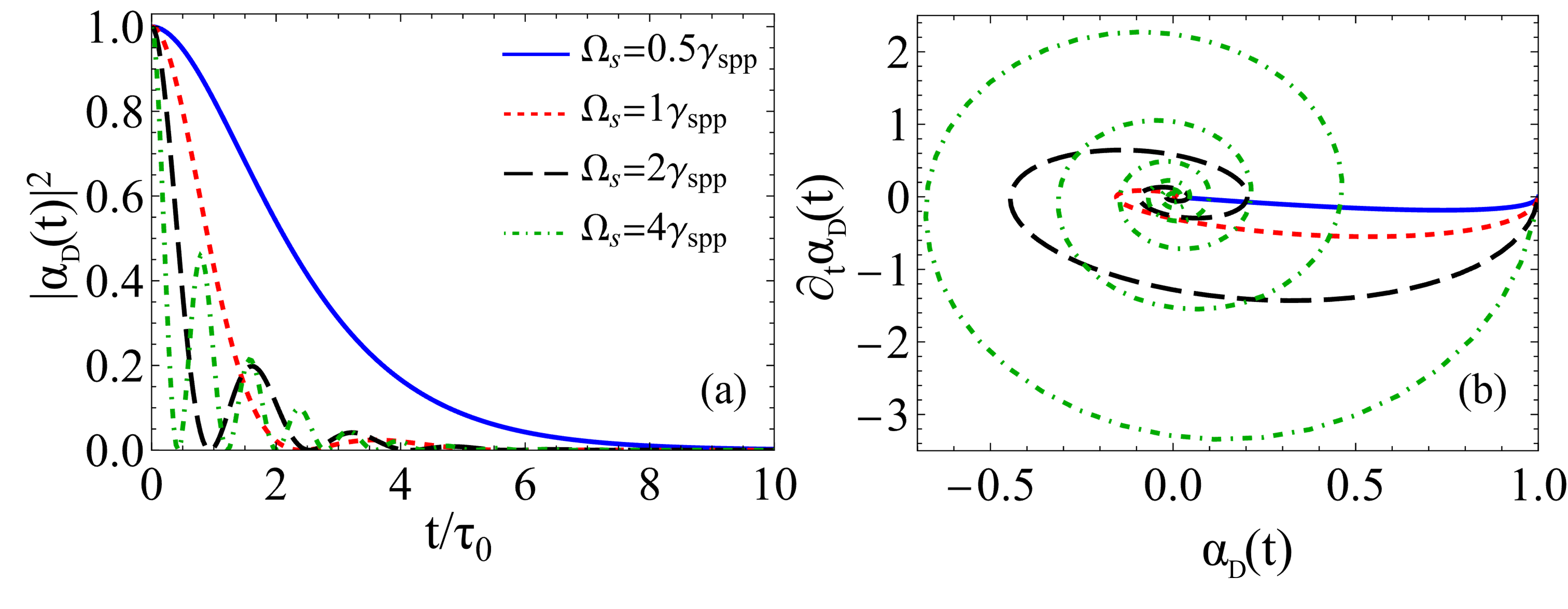}
    \caption{Panel~(a) shows the temporal evolution of $\alpha_\text{D}(t)$ governed by Eq.~\eqref{Eq:Dynamic}, corresponding to different values of effective frequency $\Omega_\text{s}/\gamma_\text{spp}$. Panel~(b) is the HPP evolution in the phase space~$(\alpha_\text{D}(t),\partial_t\alpha_\text{D}(t))$; the circulation signifies the non-Markovian SPP evolution for different $\Omega_\text{s}/\gamma_\text{spp}$. We have used $v_\text{g}=0.1\mathcal{L}\gamma_\text{spp}$, $\mathcal{L}\approx100~\mu$m and defined $\tau_0=\gamma_\text{spp}^{-1}$. See the text for other parameter values.}
    \label{fig:Temporal_Dynamics}
\end{figure}

\emph{Temporal dynamics}- We now insert these numerical values into Eq.~\eqref{Eq:Dynamic}, assume $\alpha_\text{D}(0)=1$ as the initial condition, and solve Eq.~\eqref{Eq:Dynamic} numerically for $\Omega_\text{s}\in[0.5\gamma_\text{spp},4\gamma_\text{spp}]$. The hybridized state dynamics exhibit a pure decay for $\Omega_\text{s}=0.5\gamma_\text{spp}$~(blue solid line in Fig.~\ref{fig:Temporal_Dynamics}(a)), a critical oscillatory decay for $\Omega_\text{s}=1\gamma_\text{spp}$~(red dotted line in Fig.~\ref{fig:Temporal_Dynamics}(a)), and pure oscillatory decay for $\Omega_\text{s}>\gamma_\text{spp}$, indicating the non-Markovian nature of this interaction. The observed oscillations are related to the quantum vacuum effect modified by the plasmonic density of states. This effect constitutes the most important concept of our work. Furthermore, our investigations on the phase-space evolution~(characterized by $(\alpha_\text{D}(t),\partial_t\alpha_\text{D}(t))$) justify the existence of logarithmic spirals, whose number of circulations is given implicitly by $\Omega_\text{s}/\gamma_\text{spp}$. Since we assume $\gamma_\text{D}=0$ and consider only the surface-plasmon loss, the system always decays to a fixed attractor point $$(\alpha_\text{D}(t),\dot{\alpha}_\text{D}(t))=(0,0),$$ regardless of $\Omega_\text{s}$, as shown in Fig.~\ref{fig:Temporal_Dynamics}(b).     

\begin{figure}
    \centering
    \includegraphics[width=1\linewidth]{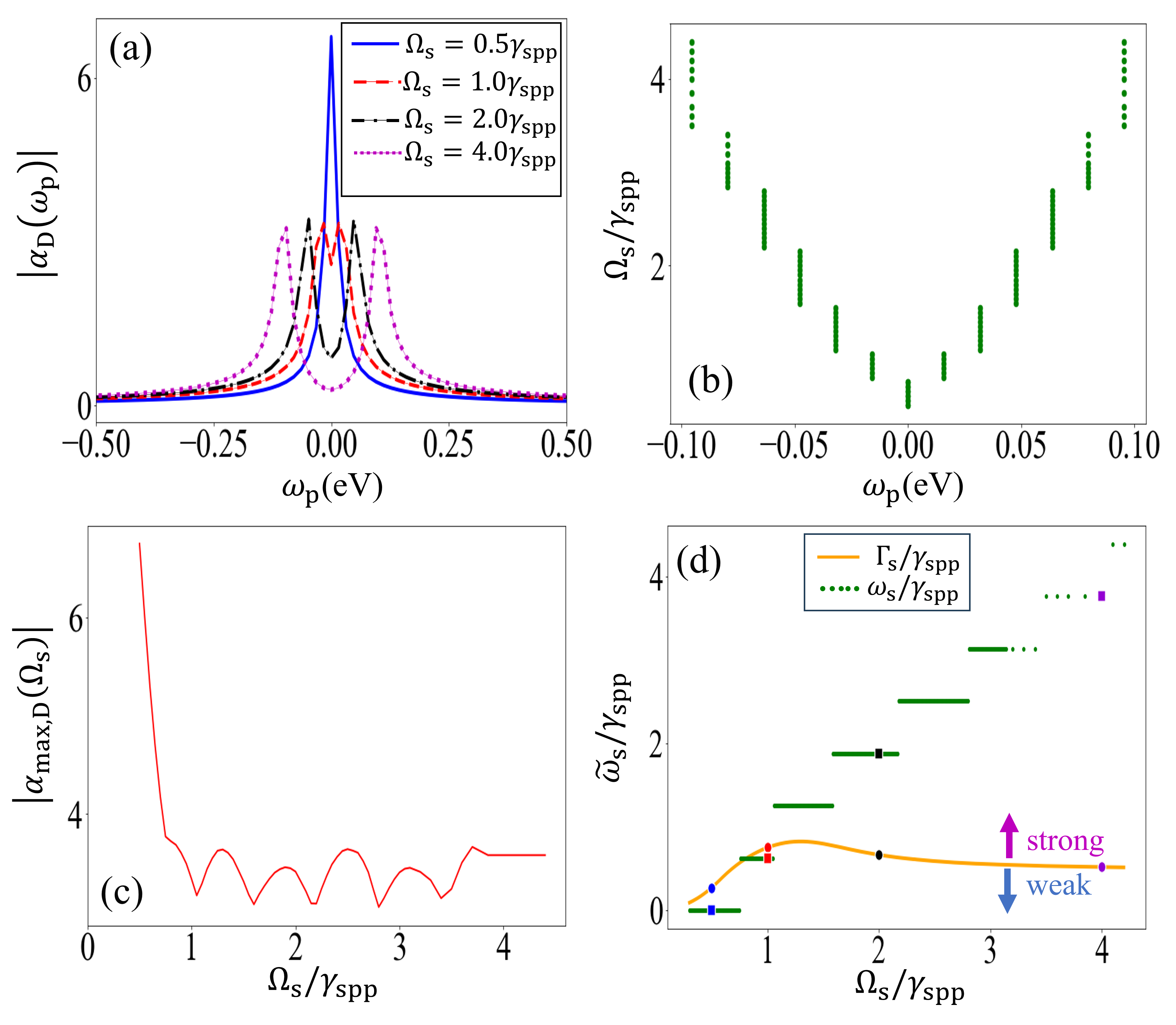}
    \caption{HPP’s spectral evolution. (a) $|\alpha_\text{D}(\omega_\text{p})|$~($\omega_\text{p}$ denoting the Fourier component) obtained via Fourier transformation of Eq.~\eqref{Eq:Dynamic} for different values of $\Omega_\text{s}$. The transition from single- to double-peaks due to normal-mode splitting is observed. (b) Position of the $\alpha_{\text{max},\text{D}}$~(the maxima of the HPP emission are shown; the corresponding amplitudes are omitted for better visibility) in the $\Omega_\text{s}-\omega_\text{p}$ plane shows a superlinear dependence of $\Omega_\text{s}$ on $\omega_\text{p}$ for $\Omega_\text{s}>\gamma_\text{spp}$, i.e., in the strong-coupling regime. Panel (c) represents the evolution of the maxima, namely, $|\alpha_{\text{max},\text{D}}|$ for different $\Omega_\text{s}$; $\alpha_{\text{max},\text{D}}$ decays exponentially by gradually increasing $\Omega_\text{s}/\gamma_\text{spp}$ but exhibits oscillations for higher values. (d) HPP’s collective oscillation $\omega_\text{s}$ and long-time decay $\Gamma_\text{s}$ for various $\Omega_\text{s}$. Blue circles and squares denote $\tilde{\omega}\text{s}$ for $\Omega_\text{s}=0.5\gamma_\text{spp}$, red for $\Omega_\text{s}=1\gamma_\text{spp}$, black for $\Omega_\text{s}=2\gamma_\text{spp}$, and violet for $\Omega_\text{s}=4\gamma_\text{spp}$. Parameters are the same as in Fig.~\ref{fig:Temporal_Dynamics}.}
    \label{fig:Spectral_Dynamcis}
\end{figure}

\emph{Spectral dynamics}- Next, we investigate the spectral evolution of the hybridized-state amplitude $\alpha_\text{D}(\omega)$ for spectral components $\omega_\text{p}\in[-\omega_\infty,+\omega_\infty]$, where $\omega_\infty$ scales with the largest frequency transition of the system. We use a discrete fast Fourier transformation
\begin{equation*}
    \alpha_\text{D}(\omega)=T_\infty^{-1}\int_0^{T_\infty}~dt' \alpha_\text{D}(t')\exp\{\text{i}\omega_\text{p}t'\},
\end{equation*}
and assume the normalization condition $$\int_{-\omega_\infty}^{\omega_\infty}d\omega_\text{p}\alpha_\text{D}(\omega_\text{p})=1,$$ and with $T_\infty=10\tau_0$ being a sufficiently long time. Our investigations indicate that for $\Omega_\text{s}<\gamma_\text{spp}$, the hybridized state irreversibly decays to the ground state, and we observe a single Lorentzian lineshape~(blue solid lines in Fig.~\ref{fig:Spectral_Dynamcis}(a)). In contrast, for $\Omega_\text{s}\geq\gamma_\text{spp}$, the coupling strengths exceed the total loss, and we observe oscillatory energy exchanges between the hybridized excited and ground states.

The spectrum exhibits normal-mode splitting~\cite{doi:10.1126/science.1244324}, as indicated by dashed red, dotted-dashed black, and dotted violet lines in Fig.~\ref{fig:Spectral_Dynamcis}(a). Although normal-mode splitting can be observed in classical systems, such as in the spectral correlations of coupled oscillators, here it arises from the interaction between the QE and plasmonic vacuum fluctuations. These insights hold for large $N$; for lower $N$, each emitter interacts with the plasmon field. In this case, $\dot{\alpha}_\text{D}(t)$ must be replaced by $\dot{\alpha}_i(t)$, and the phase $\exp\{\mathrm{i}\bm{k}\cdot(\bm{r}_i-\bm{r}_j)\}$ should be included in the dynamics. This term gives rise to additional sidebands via constructive and destructive interference between the hybridized states of individual emitters~\cite{10.1063/5.0217702}.      

Our findings, such as pure and oscillatory decay dynamics and normal-mode splitting, establish that the HPP evolution in the plasmonic \emph{waveguide} QED shares similarities with \emph{cavity} QED; nevertheless, there are features unique to our scheme that primarily stem from HPP’s non-Markovian nature. To articulate this distinction, we choose $\Omega_\text{s}$ as a tunable system parameter and calculate the maximum values of $\alpha_\text{D}(\omega)$~($|\alpha_\text{max,D}(\Omega_\text{s})|$). We represent the positions of these maxima as green sticks in Fig.~\ref{fig:Spectral_Dynamcis}(b). We observe an enhancement in the absorption-doublet splitting~(spectral separation between green dots in Fig.~\ref{fig:Spectral_Dynamcis}(b)) as $\Omega_\text{s}$ increases, confirming that an increase in effective frequency yields an enhancement in the normal-mode splitting. Interestingly, this dependency is superlinear $\Omega_\text{s}\propto\omega_\text{p}^{1+\epsilon}$~(for $0<\epsilon<1$), as shown in Fig.~\ref{fig:Spectral_Dynamcis}(b).

Furthermore, we reveal that $\alpha_{\text{max},\text{D}}$ evolves differently for various $\Omega_\text{s}$; it exhibits a sharp decay for weaker effective frequencies $\Omega_\text{s}/\gamma_\text{spp}\in[0,1]$, while for stronger hybridization $\Omega_\text{s}/\gamma_\text{spp}\in[1,4]$ the mode amplitude $\alpha_{\text{max},\text{D}}$ displays oscillations, as shown in Fig.~\ref{fig:Spectral_Dynamcis}(c). The physical reasoning behind the spectral evolution is as follows: for lower $\Omega_\text{s}$~[ $\Omega_\text{s}/\gamma_\text{spp}<1$], no sustained energy exchange occurs between the TDS and the surface plasmons at long times. Consequently, energy does not dissipate through metallic loss, resulting in a larger spectral amplitude of the HPP. For a larger effective frequency~[ $\Omega_\text{s}/\gamma_\text{spp}>1$], part of the HPP excitation is dissipated via Ohmic losses in the metallic layer, leading to a reduction in the HPP amplitude.

We also find another noticeable distinction from cavity QED by analyzing the spectral evolution of the HPP excitation. To gain a deeper insight, we compute the HPP collective frequency $\omega_\text{s}$ and long-time decay $\Gamma_\text{s}$ from Eq.~\eqref{Eq:Dynamic}, define $\tilde{\omega}_\text{s}=\omega_\text{s}+\text{i}\Gamma_\text{s}$ as the HPP excitation frequency, and evaluate $\tilde{\omega}_\text{s}$ for various $\Omega_\text{s}$~(as in Fig.~\ref{fig:Spectral_Dynamcis}(d)). We confirm that the comparison between $\omega_s$ and $\Gamma_s$ provides a clear distinction between weak- and strong-coupling regimes similar to those observed in cavity-QED. Nevertheless, dissimilarities in $\omega_\text{s}$ and $\Gamma_\text{s}$ emerge in the frequency plane characterized by $(\Gamma_\text{s},\omega_\text{s})$, where we identify three long-time decay regimes: (i) for $0.5\gamma_\text{spp}<\Omega_\text{s}<\gamma_\text{spp}$, $\Gamma_\text{s}$ increases, (ii) for $\Omega_\text{s}\approx\gamma_\text{spp}$, $\Gamma_\text{s}$ reaches a maximum, and, (iii) for $\Omega_\text{s}>\gamma_\text{spp}$, $\Gamma_\text{s}$ decreases while asymptotically approaches $0.5\gamma_\text{spp}$ for large $\Omega_\text{s}$~(see Fig.~\ref{fig:Spectral_Dynamcis}(d)). This situation differs fundamentally from the conventional QED systems and stems from the fact that the linear dispersion and non-Markovianity of the HPP evolution are encoded in these quantities through the memory kernel in Eq.~\eqref{Eq:Dynamic}.

\begin{figure}
    \centering
    \includegraphics[width=1\linewidth]{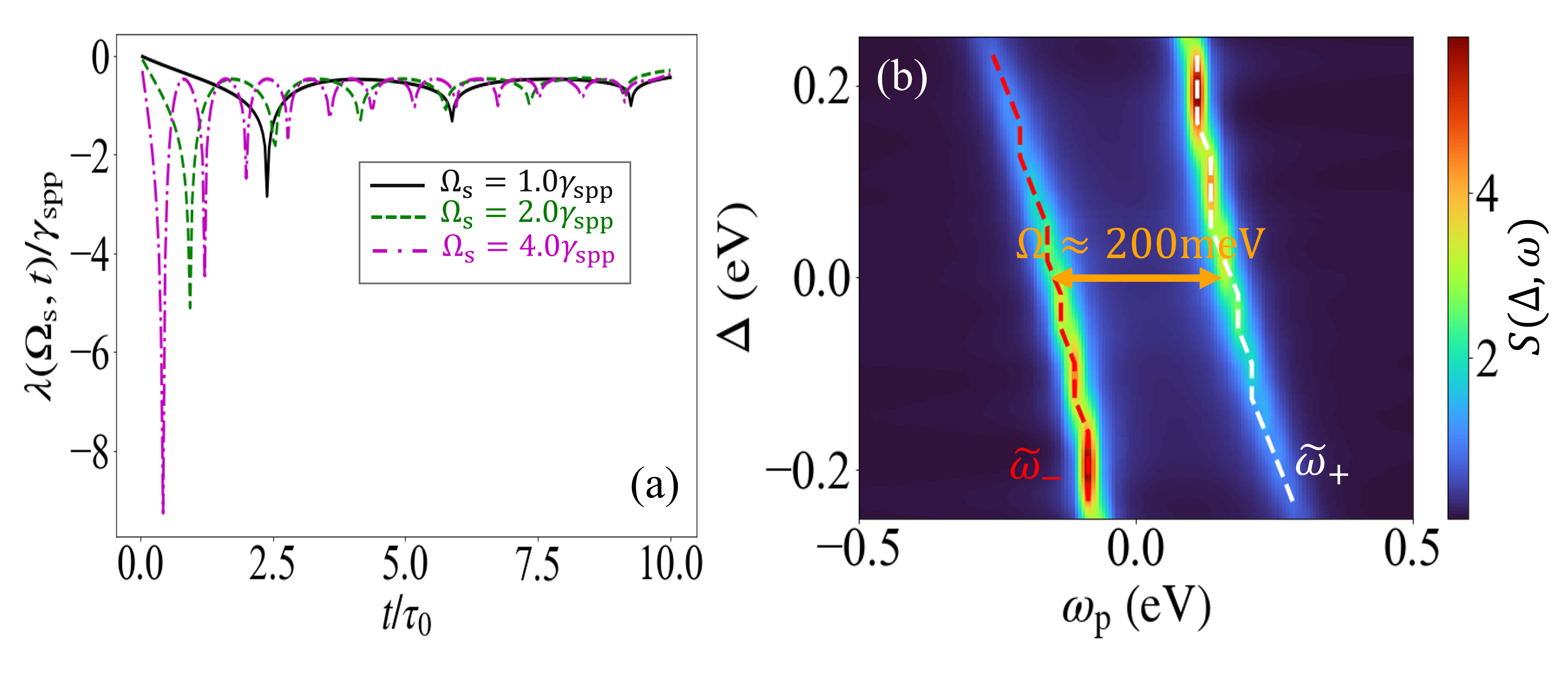}
    \caption{(a) Dynamics of Lyapunov exponent $\lambda$ for various ranges of effective Rabi frequency, showing stable three decay regimes: early-time fast, transient-time oscillatory, and long-time classic. Panel (b) displays the emission spectrum for different detunings $\Delta:=\omega_\text{eg}-\omega_\text{spp}$ and Fourier frequency components $\omega_\text{p}$ for $\Omega_\text{s}=4\gamma_\text{spp}$. We characterize the evolution of the positive branch~($\tilde{\omega}_+$, dashed white curve) and negative branch ($\tilde{\omega}_-$, dashed red curve), and obtain with peak splitting up to $\Omega=200$~meV. Parameters are the same as Fig.~\ref{fig:Temporal_Dynamics}.}
    \label{fig:Quantum_Statistics}
\end{figure}

The existence of $\Gamma_\text{s}$ and $\omega_\text{s}$ implies that Eq.~\eqref{Eq:Dynamic} admits a classical damped harmonic representation of the form
\begin{equation}
    \ddot{\alpha}_\text{D}(t)+\Gamma_\text{s}\dot{\alpha}_\text{D}(t)+\omega_\text{s}^2\alpha_\text{D}(t)=0,
    \label{Eq:General_Dynamics_SPP}
\end{equation}
where we obtain the coefficients using a nonlinear interpolation technique. The mapping between Eq.~\eqref{Eq:Dynamic} and Eq.~\eqref{Eq:General_Dynamics_SPP} is heuristic; nevertheless, they indicate that the HPP dynamics can also be understood as a time-scaling problem. Specifically, on the basis of Eq.~\eqref{Eq:Dynamic}, and by defining memory time as $\tau_\text{mem}:=\sqrt{2}\mathcal{L}/v_\text{g}$ and coupling time as $\tau_\text{cou}:=\Omega_\text{s}^{-1/2}$, various coupling regimes can be achieved through the interplay between $\tau_\text{mem}$, $\tau_\text{cou}$ and $\tau_0$. For $v_\text{g} = 0.1\gamma_\text{spp}\mathcal{L}$ we obtain 
\begin{equation*}
   \tau_\text{mem}\gg\tau_0,\tau_\text{cou}, 
\end{equation*}
thereby, various regimes can then be achieved by comparing $\tau_0$ and $\tau_\text{cou}$: weak coupling occurs for $\tau_\text{cou}>\tau_0$ whereas strong coupling is expected for $\tau_\text{cou}<\tau_0$. In the harmonic-oscillator representation~(namely, using Eq.~\eqref{Eq:General_Dynamics_SPP}), by defining
\begin{equation*}
    \tau_\text{cou}'=\omega_\text{s}^{-1},\quad \tau_0':=\Gamma_\text{s}^{-1},
\end{equation*}
we obtain weak coupling for $\tau_\text{cou}'>2\tau_0'$, whereas strong coupling emerges when $\tau_\text{cou}'<2\tau_0'$~(see appendix \ref{Sec:Physical_Range} for further discussion on time scales).  

\subsection{\label{Sec:Lyapunov_Exponent} Lyapunov exponent and decay regimes}
We note that Eq.~\eqref{Eq:General_Dynamics_SPP}, despite offering a sophisticated description of the system, has a purely classical structure; hence, the quantum-like nature of the HPP evolution decay appears to be overlooked and unverifiable within this framework. To uncover and analyze this hidden quantum feature~(termed instantaneous decay~\cite{PhysRevLett.131.193603,PhysRevResearch.6.033243}), we introduce and calculate the finite-time Lyapunov exponent $\lambda$. To this end, we exploit the numerical solution of Eq.~\eqref{Eq:Dynamic} or Eq.~\eqref{Eq:General_Dynamics_SPP}, and define
\begin{equation*}
    \delta\alpha_{\Omega_\text{s}}(t)=\alpha_\text{D}(t+\delta t)-\alpha_\text{D}(t),
\end{equation*}
with $\delta t =10^{-6}\tau_0$ the small perturbation parameter. We then achieve the finite-time Lyapunov exponent as~\cite{PhysRevE.89.032918}
\begin{equation}
    \lambda(\Omega_\text{s},T_\delta) = \lim_{\delta t\to T_\delta}t^{-1}\ln(\left|\delta\alpha_{\Omega_\text{s}}(t+\delta t)/\delta\alpha_{\Omega_\text{s}}(t)\right|).
    \label{Eq:Lyapunov}
\end{equation}
Here, we discretize the temporal window $[0,10\tau_0]$ into $2^{10}$ temporal grids~($T_\delta$), and evaluate $\lambda$ for each of them.  

Lyapunov exponent analysis uncovers novel features regarding HPP evolution, as indicated in Fig.~\ref{fig:Quantum_Statistics}(a). First, regardless of $\Omega_\text{s}$, we obtain $\lambda(\Omega_\text{s},T_\delta)<0$ over the entire temporal window, demonstrating that the dynamics are stable and eventually relax to an attractor point $(0,0)$~(decay to the ground state). Therefore, for sufficiently long times, $\lambda(\Omega_\text{s})$ asymptotically approaches $\Gamma_\text{s}\approx 0.5\gamma_\text{spp}$. Next, at early times, we observe a fast HPP decay that scales~(approximately) as 
\begin{equation*}
    \lambda(\Omega_\text{s},T_\delta)\sim\Omega_\text{s}^2\propto N^2.
\end{equation*}
This fast decay can have a counterpart in classical non-Markovian systems, for instance, in phase-locked coupled oscillators; however, its quantum origin lies in the single-photon excitation and intrinsic quantum vacuum effects. In particular, we note that, since the system is prepared in the TDS at $t=0$, the rapid decay observed at early times $t<\tau_\text{cou}$ corresponds to superradiant emission. Here, we mention that any deviation from $\Omega_\text{s}^2$ scaling originates in the lossy and non-Markovian nature of the HPP field. 

Interestingly, we obtain an oscillatory evolution at transient times, which implies energy exchange between the hybridized and ground states. This transient regime reveals strong coupling between the TDS~(where maxima at later times are reduced due to the HPP decay~[see Fig.~\ref{fig:Quantum_Statistics}(a)]), and the surface-plasmon field~(whose decay is classical and uniform in time); hence, at these times, the instantaneous decay represents non-classical behavior. These insights indicate that both long-time and instantaneous decays are essential for a complete description of the HPP dynamics; here, the instantaneous decay acts as a snapshot of the dynamics on small temporal grids. We finally note that the instantaneous decay has already been investigated~\cite{PhysRevLett.131.193603,PhysRevResearch.6.033243} using
\begin{equation}
    \Gamma_\text{inst}(t)=-\frac{d}{dt}\left(\ln\{P_\text{e}(t)\}\right)
    \label{Eq:Specific_Decay}
\end{equation}
for $$P_\text{e}(t)=\sum_{i}|\alpha_{\text{e},i}(t)|^2,$$ where $\alpha_{\text{e},i}(t)$ are the $i$th emitter’s excited state. However, Eq.~\eqref{Eq:Specific_Decay}
can be problematic for oscillatory decay~(when $P_\text{e}(t)\to0$) or when the system and bath coherently exchange energy through the Rabi oscillation.      

\subsection{\label{Sec:Discussion}Discussion}
Following the temporal and spectral characterization of the HPP, we now provide a deeper insight into the spectral features of our plasmonic WQED. To this end, we define $\Delta = \omega_\text{spp}-\omega_\text{eg}$ as detuning and consider it as a control parameter, assume $\Omega_\text{s}=4\gamma_\text{spp}$, and calculate the emission spectrum using 
\begin{equation}
    S_\text{em}(\Delta,\omega)\propto \Re\Big\{\sum_i \int_{0}^{T_\infty}dt \bra{\Psi(0)} \sigma_+^{(i)} \sigma_-^{(i)} \ket{\Psi(t)}\text{e}^{\text{i}\omega t}\Big\}.
    \label{Eq:Emission}
\end{equation}
We observe an \emph{anticrossing} of the peak doublet, similar to the HPP’s spectral evolution~(see Fig.~\ref{fig:Spectral_Dynamcis}(a) and Fig.~\ref{fig:Quantum_Statistics}(b)). Specifically, we describe the spectral evolution of the $S_\text{em}(\Delta,\omega)$ peaks in the $(\Delta,\omega)$ plane identified as the negative $\tilde{\omega}_-$ and positive $\tilde{\omega}_+$ branches of the hybridized states~(see Fig.~\ref{fig:Quantum_Statistics}(b)), and confirm the emergence of an anticrossing similar to that seen in the cavity QED. 

Despite these similarities between our WQED and conventional cavity QED~\cite{RevModPhys.94.041003}, the underlying physics of the anticrossing differs. Indeed, the peak-splitting in $\tilde{\omega}_\pm$ emerges within the \textit{interaction} picture and is directly associated with the HPP oscillations~(reflecting the HPP’s emission–re-emission) in the non-Markovian framework, whereas in the cavity QED, the hybridization appears in the \emph{bare-state} frame and originates from the coupling between a single-cavity mode and QEs. 

We have also tuned various control parameters, such as $v_\text{g}$, $k$, and $\Omega_\text{s}/\gamma_\text{spp}\in\{1,2,3\}$, and find similar features in the evolution of $\omega_\pm$. This anti-crossing appears to be a generic feature associated with the structured quantum vacuum and the resulting non-Markovian light–matter interaction. We therefore predict that this anti-crossing feature may also be realized in WQED setups, such as emitter–fiber systems~\cite{PhysRevLett.130.163602}, provided that the reservoir is sufficiently structured, the TDS decay into free space is negligible, and the dynamics exhibit oscillatory decay.   

\section{\label{Sec:Conclusion_Outlook} Conclusion and outlook}

\subsection{Conclusion}
To sum up, we explore the collective-light–collective-matter interaction in nanoscopic WQED, comprising a TDS situated on top of a metallic layer. Our analysis of TDS does not include a fully microscopic, many-body treatment, but instead describes a collective, non-Markovian evolution. In the interaction picture, this coupling gives rise to HPP excitation and propagation, whose dynamics can be described by combining the macroscopic QED framework with Fourier optics of the surface-plasmon field. By employing the momentum-matching condition and the surface-plasmon dispersion relation and within the Schr\"odinger framework, we obtain and solve an integrodifferential equation governing $\alpha_\text{D}(t)$ to achieve HPP’s temporal and spectral evolution. Specifically, we introduce $\Omega_\text{s}$ as a well-defined parameter, for which HPP dynamics show pure decay~(weak coupling) for $\Omega_\text{s}<\gamma_\text{spp}$, whereas we observe oscillatory decay~(strong coupling) for $\Omega_\text{s}>\gamma_\text{spp}$. In phase space, the HPP evolves as logarithmic spirals that~(independent of system parameters) stably decay to an attractor point, while $\Omega_\text{s}$ determines the number of circulations. Spectral analysis of the hybrid state reveals normal-mode splitting, indicating a strong-coupling regime, originating from the vacuum-field Rabi oscillation. We find that this splitting is adjustable by tuning $\Omega_\text{s}$. The peak position $|\alpha_\text{max,D}|$ in the $(\omega_\text{p},\Omega_\text{s})$ plane exhibits a superlinear dependence on $\Omega_\text{s}$. 

We then show that the HPP dynamics admit a damped-harmonic oscillator representation, describable via a collective frequency and a long-time decay; this interpretation confirms the existence of weak- and strong-coupling regimes similar to cavity QED. In addition to long-time decay, we also exploit finite-time Lyapunov-exponent analysis, which enables us to characterize instantaneous decay, exhibiting three distinct regimes: fast quantum-like decay~($\lambda(\Omega_\text{s},T_\delta)\propto \Omega_\text{s}^2$), non-classical oscillatory decay, and classical long-time decay. We show that the HPP’s non-Markovian evolution in plasmonic \emph{waveguide} QED displays anticrossing~(even up to $\Omega\approx200$~meV) in the emission spectrum and exhibits hybridized frequency $\tilde{\omega}_\pm$ evolutions, similar to those observed in \emph{cavity} QED, but with a different anticrossing mechanism: ours occurs in the interaction picture, whereas the splitting in cavity QED is based on coupled-mode and bare-state analysis.                               
\subsection{Outlook}
Apart from the relaxation mechanisms, field inhomogeneities, and many-body effects discussed in this letter, we outline two directions for future work. The first direction concerns dissimilarities between plasmonic WQED and cavity QED; our analysis of long-time decay in Fig.~\ref{fig:Spectral_Dynamcis}(d) shows anomalous behavior, for which the HPP’s collective oscillation becomes independent of loss for large $\Omega_\text{s}$. We suggest that this behavior is related to the non-Markovian nature of the interaction; however, this spectral evolution requires further analysis that can be assumed as future work~(see also \S~\ref{Sec:Dissimilar_Cavity_QED} and Fig.~\ref{Fig:Strong_Weak_Coupling} of the appendix). We also suggest investigating the ultra-strong coupling regime~\cite{frisk2019ultrastrong,PhysRevResearch.5.033002,PhysRevLett.134.063602} in this plasmonic WQED platform. This regime can be achieved by increase effective frequency to $\Omega_\text{s}/\omega_\text{eg}=0.1$. Increasing the light–matter coupling strength can provide the conditions required for the emergence of virtual particles and other intriguing features of the ultrastrong-coupling regime~(see also appendix \ref{Sec:Master_Equation_USC_Regime} for more details).

\appendix
\begin{widetext}
\section{\label{Derivation_Equations} Quantitative details towards Derivation of Eqs.~(10) and (12)}
In this section, we provide detailed qualitative and quantitative steps leading to the derivation of Eqs.~\eqref{Eq:Wavefunction_Momentum} and \eqref{Eq:Dynamic} of the main text. First, in \S~\ref{Approach:Schrodinger}, we elucidate the Schr\"odinger-equation approach. Then, in \S~\ref{Sec:Challenge_Dynamics}, we derive a master equation capable of describing HPP dynamics and discuss the possible challenges associated with applying the master-equation formalism to our plasmonic WQED.

\subsection{\label{Approach:Schrodinger} Schr\"odinger approach}
We now derive and analyze the dynamical evolution of the hybridized-state transition amplitude $\alpha_\text{D}(t)$, which is directly related to the evolution of the HPP for $t>0$. We work in the rotating frame defined with respect to the bare quantum-emitter (QE) Hamiltonian $H_\text{e}$ and the surface-plasmon Hamiltonian $H_\text{f}$. In this frame, the interaction Hamiltonian is given by Eq.~\eqref{Eq:Hamiltonian_Int} in the main text, and the state vector of the hybridized system is expressed as in Eq.~\eqref{Eq:Wavefunction}, where $\ket{\mathcal{G}}$ denotes the collective ground state of the quantum emitters, and $\ket{0_{\bm{r}',\tilde{\omega}}} \equiv \ket{0}$ represents the vacuum state of the surface-plasmon field at position $\bm{r}'$ and frequency $\tilde{\omega}$. The state $\ket{\Psi(N|k)}$ corresponds to the maximally symmetric superposition state~(TDS), while $\ket{\Psi_{l\perp}}$ denotes the sum over all other Dicke states orthogonal to the TDS. 

In this work, we follow the conventional definition of the TDS for an ensemble of $N$ quantum emitters
\begin{equation}
    \ket{\Psi_\text{Dicke}}=\frac{1}{\sqrt{N}}\sum_{i=1}^N e^{\text{i}\bm{k}\cdot\bm{r}_i}\ket{e_i}\underset{i\neq j}\otimes\ket{g_j},
    \label{SEq:Dicke_tradition}
\end{equation}
and use \eqref{SEq:Dicke_tradition} to express the hybridized state wavefunction of the system as $\ket{\Psi_\text{D}(N|\bm{k})}:=\ket{\Psi_\text{Dicke}}\otimes\ket{0}$, namely
\begin{equation}
    \ket{\Psi_\text{D}(N|\bm{k})}=\left(\frac{1}{\sqrt{N}}\sum_{i=1}^N \text{e}^{\text{i}\bm{k}\cdot\bm{r}_i}\ket{e_i}\underset{i\neq j}\otimes\ket{g_j}\right)\otimes\ket{0}.
    \label{SEq:Dicke}
\end{equation}
We then introduce the following abbreviated notation for the TDS~\cite{doi:10.1021/acsphotonics.9b00193}:
\begin{equation}
\ket{\Psi_\text{D}(N|k)}=\frac{1}{\sqrt{N}}\sum_{i=1}^N
\mathrm{e}^{\mathrm{i}\bm{k}\cdot\bm{r}_i}
\ket{e_i}\underset{i\neq j}{\otimes}\ket{g_j,{0}},
\end{equation}
where $\ket{g_j,{0}} := \ket{g_j}\otimes\ket{{0}}$ denotes a shorthand notation. This definition yields Eq.~\eqref{Eq:Timed_Dicke_State} in the main text. Owing to the directionality of the Dicke states, including the TDS, the resulting HPP field propagates in the same direction as the photon with wavenumber $\bm{k}$ that prepared the TDS. Using the resultant directionality as the unique feature of the TDS, we study the dynamical evolution of the system by inserting Eq.~\eqref{Eq:Hamiltonian_Int} and Eq.~\eqref{Eq:Wavefunction} into the Schr\"odinger equation
\begin{equation}
    i\hbar \ket{\dot{\Psi}(t)}=H_\text{int} \ket{\Psi(t)}.
\end{equation}

We obtain the temporal evolution of the transition amplitudes in Eq.~\eqref{Eq:Wavefunction} by introducing $\ket{\Psi_\text{G}}:=\ket{\mathcal{G}}\otimes\ket{0}$ as the total ground state and defining $\mathcal{C}:=1/\sqrt{\hbar\pi\varepsilon_0}$ and $\varepsilon_\text{i}(\bm{r}',\tilde{\omega}):=\Im[\varepsilon_\text{m}(\bm{r}',\tilde{\omega})]$, as
\begin{align}
\dot{\xi}(\mathbf r',t;\tilde{\omega}) ={}& -\mathcal{C}\sum_{l=\mathrm{D},0\perp}^{(N-1)\perp}\sum_{i=1}^{N}\left(\frac{\tilde{\omega}}{\mathrm{c}}\right)^2\sqrt{\varepsilon_{\mathrm i}(\mathbf r',\tilde{\omega})}\bra{\Psi_{\mathrm G}}\sigma_-^{(i)}\hat{\mathbf f}^{\dagger}(\mathbf r',\omega)\cdot\mathbf G^*(\mathbf r',\mathbf r_i;\tilde{\omega})\cdot\mathbf p_i\ket{\Psi_l(\mathbf k)}\alpha_l(t)e^{i(\tilde{\omega}-\omega_{\mathrm{eg}})t},
\label{SEq:Ground_State}
\\
\dot{\alpha}_{\mathrm D}(t) ={}& \mathcal{C}\sum_{i=1}^{N}\int d\tilde{\omega}\left(\frac{\tilde{\omega}}{\mathrm{c}}\right)^2\int d\mathbf r'\sqrt{\varepsilon_{\mathrm i}(\mathbf r',\tilde{\omega})}\bra{\Psi_{\mathrm D}(N|\mathbf k)}\sigma_+^{(i)}\mathbf p_i\cdot\mathbf G(\mathbf r_i,\mathbf r';\tilde{\omega})\cdot\hat{\mathbf f}(\mathbf r',\omega)\ket{\Psi_{\mathrm G}}\xi(\mathbf r',t;\tilde{\omega})e^{-i(\tilde{\omega}-\omega_{\mathrm{eg}})t},
\label{SEq:Dicke_Superposition}
\\
\dot{\alpha}_{l\perp}(t) ={}& \mathcal{C}\sum_{i=1}^{N}\int d\tilde{\omega}\left(\frac{\tilde{\omega}}{\mathrm{c}}\right)^2\int d\mathbf r'\sqrt{\varepsilon_{\mathrm i}(\mathbf r',\tilde{\omega})}\bra{\Psi_{l\perp}(\mathbf k)}\sigma_+^{(i)}\mathbf p_i\cdot\mathbf G(\mathbf r_i,\mathbf r';\tilde{\omega})\cdot\hat{\mathbf f}(\mathbf r',\omega)\ket{\Psi_{\mathrm G}}\xi(\mathbf r',t;\tilde{\omega})e^{-i(\tilde{\omega}-\omega_{\mathrm{eg}})t}.
\label{SEq:Dicke_Superposition_Middle}
\end{align}
To achieve the transition amplitude of the hybridized state $\ket{\Psi_\text{D}(N|\bm{k})}$~(namely $\alpha_\text{D}(t)$), we eliminate the ground-state evolution $\xi(\bm{r}',t;\tilde{\omega})$ by performing a time-integration of Eq.~\eqref{SEq:Ground_State} and then insert the resulting expression into Eq.~\eqref{SEq:Dicke_Superposition} and \eqref{SEq:Dicke_Superposition_Middle}. We then utilize the orthogonality condition of the Dicke states, that is, $\braket{\Psi_\text{D}(N|\bm{k})|\Psi_{l\perp}(\bm{k})}=0$. During the time integration, we obtain the following product of the average values
\begin{equation*}
    \bra{\Psi_\text{D}(N|\bm{k})}\sigma_+^{(i)}\bm{p}_i\cdot\bm{G}(\bm{r}_i,\bm{r}';\tilde{\omega})\cdot\hat{\bm{f}}(\bm{r}',\omega)\ket{\Psi_\text{G}}\bra{\Psi_\text{G}}\sigma_-^{(j)}\hat{\bm{f}}^{\dagger}(\bm{r}',\omega)\cdot\bm{G}^*(\bm{r}',\bm{r}_j;\tilde{\omega})\cdot\bm{p}_j\ket{\Psi_l(\bm{k})},
\end{equation*}
which can be simplified, in the single excitation regime, to
\begin{equation}
    \bra{\Psi_\text{D}(N|\bm{k})}\sigma_+^{(i)}\sigma_-^{(j)}\ket{\Psi_l(\bm{k})}\bm{p}_i\cdot\bm{G}(\bm{r}_i,\bm{r}';\tilde{\omega})\cdot\bm{G}^*(\bm{r}',\bm{r}_j;\tilde{\omega})\cdot\bm{p}_j.
    \label{Eq:Green_multilication}
\end{equation}
The spatial integration of the \eqref{Eq:Green_multilication} is proportional to $\alpha_\text{D}(t)$, which can be accurately characterized by introducing the following abbreviation
\begin{equation}
    R_{ij}=\bm{p}_i\cdot\left\{\left(\frac{\tilde{\omega}}{\text{c}}\right)^2\int~d^3\bm{r}'\varepsilon_\text{i}(\bm{r}',\tilde{\omega})\bm{G}(\bm{r}_i,\bm{r}';\tilde{\omega})\cdot\bm{G}^*(\bm{r}',\bm{r}_j;\tilde{\omega})\right\}\cdot\bm{p}_j.
    \label{Eq:Dummy_Variable}
\end{equation}
Then we leverage Eq.~\eqref{Eq:Dummy_Variable} to rewrite Eq.~\eqref{SEq:Dicke_Superposition} as
\begin{equation}
\label{SEq:Dynamics_DickeIntermediate}
    \dot{\alpha}_\text{D}(t) \propto -\sum_{ij=1}^N\int_0^t d\tau\int d\tilde{\omega}\left(\frac{\tilde{\omega}}{\text{c}}\right)^{2} R_{ij} \bra{\Psi_\text{D}(N|\bm{k})}\sigma_+^{(i)}\sigma_-^{(j)}\ket{\Psi_l(\bm{k})}\alpha_\text{D}(t)\text{e}^{-i(\tilde{\omega}-\omega_\text{eg})(t-\tau)}.    
\end{equation}
The expression $\bra{\Psi_\text{D}(N|\bm{k})}\sigma_+^{(i)}\sigma_-^{(j)}\ket{\Psi_\text{D}(\bm{k})}$ in \eqref{SEq:Dynamics_DickeIntermediate} can be further simplified using Eq.~\eqref{SEq:Dicke} to
\begin{equation*}
     \bra{\Psi_\text{D}(N|\bm{k})}\sigma_+^{(i)}\sigma_-^{(j)}\ket{\Psi_\text{D}(\bm{k})}=\frac{1}{N}\text{e}^{i\bm{k}\cdot(\bm{r}_i-\bm{r}_j)}.
\end{equation*}
Furthermore, the expression $R_{ij}$ in Eq.~\eqref{Eq:Dummy_Variable}, which involves a spatial integration of the product of the Green’s function, can also be simplified using the following identity
 \begin{equation}
     \frac{\tilde{\omega}^2}{\pi\text{c}^2} \int d \bm r'  \epsilon_\text{I}(\bm{r}',\tilde{\omega}) \bm{G}(\bm{r}_j,\bm{r}';\tilde{\omega}).\bm{G}^*(\bm{r}_i,\bm{r}';\tilde{\omega})=\Im[\bm{G}(\bm{r}_j,\bm{r}_i;\tilde{\omega})].
     \label{Eq:Green_Identity}
 \end{equation}
Plugging these expressions into Eq.~(\ref{SEq:Dynamics_DickeIntermediate}), we obtain the real-space dynamics of the TDS as
\begin{equation}
    \label{SSEq:Dynamics_DickeIntermediate}
    \dot{\alpha}_\text{D}(t) =-\frac{\mathcal{C}}{N}\sum_{ij=1}^N\int_0^t d\tau\int d\tilde{\omega}\left(\frac{\tilde{\omega}}{\text{c}}\right)^{2}
    \left(\bm{p}_i\cdot\Im[\bm{G}(\bm{r}_j,\bm{r}_i;\tilde{\omega})]\cdot\bm{p}_j \right)
    \alpha_\text{D}(\tau)\text{e}^{\text{i}\left[\bm{k}\cdot(\bm{r}_i-\bm{r}_j)-(\tilde{\omega}-\omega_\text{eg})(t-\tau)\right]}.
\end{equation}
To apply Eq.~\eqref{SSEq:Dynamics_DickeIntermediate} for our plasmonic WQED, the linear dispersion of the metal-dielectric interface, namely, the dependency of $\tilde{\omega}$ on the in-plane momentum $\bm{q}_\parallel$~(see \S~\ref{Sec:General_SPP} for more details) has to be taken into account. To this aim, we employ the Fourier-space representation of the Green’s function for a real wavevector $\bm{q}_\parallel$ and complex frequency $\tilde{\omega}$~\cite{PhysRevB.79.195414}. Specific to a flat plasmonic nanostructure, we can leverage the in-plane translational symmetry of the waveguide to rewrite the Green’s function in Eq.~\eqref{SSEq:Dynamics_DickeIntermediate} as 
\begin{equation}
    \bm{G}(\bm{r}_i,\bm{r}_j;\tilde{\omega})=\int\frac{d^2 \bm{q}_\parallel}{(2\pi)^2}\bm{G}(z,z';\bm{q}_\parallel,\tilde{\omega})\text{e}^{i\bm{q}_\parallel\cdot(\bm{r}_i-\bm{r}_j)}.
\end{equation}
By inserting this expression into \eqref{SSEq:Dynamics_DickeIntermediate}, and defining the emitter-emitter coupling strength as 
\begin{equation}
    \mathcal{J}_{ij}(\bm{q}_\parallel,\tilde{\omega})=\Im\left[\left(\frac{\tilde{\omega}}{\text{c}}\right)^2\bm{p}_i\cdot\bm{G}(z_i,z_j;\bm{q}_\parallel,\tilde{\omega})\cdot\bm{p}_j\right],
    \label{SEq:Emitter_Emitter}
\end{equation}
we obtain the temporal evolution of the TDS excitation amplitude as
\begin{equation}
    \dot{\alpha}_\text{D}(t) =-\mathcal{C}'\sum_{ij=1}^N\int_0^t d\tau\int d\tilde{\omega}\int\frac{d^2 \bm{q}_\parallel}{(2\pi)^2}\mathcal{J}_{ij}(\bm{q}_\parallel,\tilde{\omega})\alpha_\text{D}(\tau)\text{e}^{i\left[(\bm{k}-\bm{q}_\parallel)\cdot(\bm{r}_i-\bm{r}_j)-(\tilde{\omega}-\omega_\text{eg})(t-\tau)\right]}.
    \label{SEq:Dicke_DynamicsGeneral}
\end{equation}

\begin{figure*}
    \centering
    \includegraphics[width=0.8\linewidth]{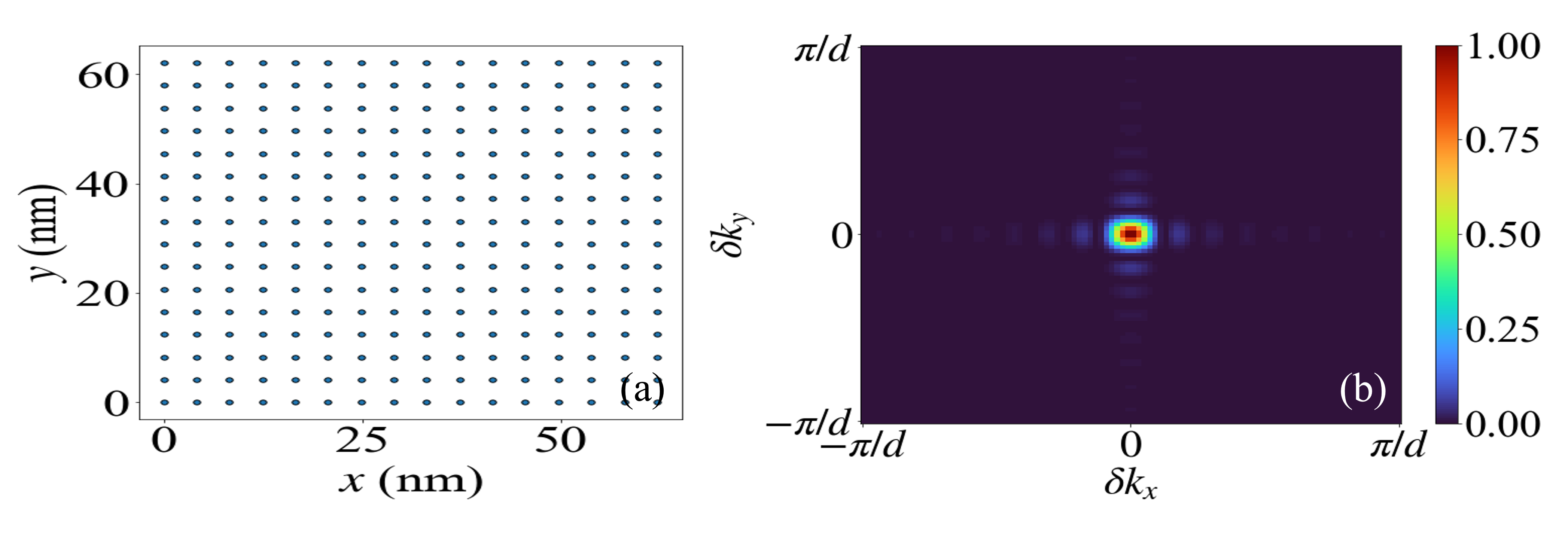}
    \caption{Momentum-matching condition of the TDS for a square lattice of the QEs: panel (a) represents the $16\times16$ square lattice of the QEs situated on top of a flat metallic layer; $x$ and $y$ axes are scaled in nanometers. Panel (b) represents $\sum_{ij=1}^{16}\exp\{\text{i}(\bm{q}_\parallel-\bm{k})\cdot(\bm{r}_i-\bm{r}_j)\}$ as a function of $\delta k_x=\bm{q}_{\parallel x}-\bm{k}_x$ and $\delta k_y=\bm{q}_{\parallel y}-\bm{k}_y$. We see a clear localization in momentum space, which guarantees the directionality of the TDS and momentum-matching condition $\bm{q}_\parallel=\bm{k}$. We have used $\bm{r}_l=d(n\bm{x}+m\bm{y})$; $l\in\{i,j\}$ and $n,m\{1,2,\ldots,16\}$ and set $d=4$~nm. See the text for more details. }
    \label{fig:SJustification_Momentum}
\end{figure*}

Eq.~\eqref{SEq:Dicke_DynamicsGeneral} represents the most general form of the dynamics of the TDS amplitude transition in our plasmonic WQED. This equation contains quantities associated with both the plasmonic waveguide and the TDS; consequently, its temporal evolution may reflect the properties of the HPP state. In particular, plasmonic waveguide properties such as the optical density of states are encoded in the TDS dynamics~[Eq.~\eqref{SEq:Dicke_DynamicsGeneral}] through the dissipative emitter-emitter coupling. In what follows, we exploit two assumptions to solve \eqref{SEq:Dicke_DynamicsGeneral}, namely, (i) QEs are distributed in the $(x,y)$ plane at the same height relative to the interaction interface (i.e., $z_i = z_j$), and possess identical polarizabilities ($\bm p_i = \bm p_j$). For this specific case, $\mathcal{J}_{ij}(\bm{q}_\parallel,\tilde{\omega})$ becomes independent of the indices $i$ and $j$, and (ii) We assume that the emitter ensemble contains a finite number of particles, $N$. The first assumption allows us to reduce the three-dimensional spatial distribution of emitters to a two-dimensional ensemble, which we model as a lattice structure, and the second assumption enables us to approximate the in-plane momentum-space distribution of emitters $\zeta(\bm{q}_\parallel,\bm{k})$ with a Gaussian distribution function around $\bm{k}$ with a finite width $\mathcal{L}$, namely,    
\begin{equation}
     \zeta(\bm{q}_\parallel,\bm{k}):=\sum_{ij=1}^N \text{e}^{\text{i} (\bm{q}_\parallel-\bm{k})\cdot(\bm{r}_i-\bm{r}_j)}\sim N^2\text{e}^{-\mathcal{L}^2(\bm{q}_\parallel-\bm{k})^2}.
     \label{SEq:Finite_number}
\end{equation}

Equation~\eqref{SEq:Finite_number} explicitly implies the directionality of the HPP, governing its temporal and spectral evolution. It further shows that, among all possible in-plane momenta $\bm{q}\parallel$, only those satisfying the momentum-matching condition $\bm{q}\parallel \approx \bm{k}$ can be efficiently excited and propagate. Qualitatively, this feature originates from the TDS, whose directionality has already been established for a large number of QEs. Specifically, in the limit of $N \to \infty$, Eq.~\eqref{SEq:Finite_number} reduces to $\zeta(\bm{q}_\parallel,\bm{k}) \to \delta(\bm{q}_\parallel - \bm{k})$, which is consistent with the directionality assumptions of the TDS~\cite{PhysRevLett.96.010501}. For lower numbers of quantum emitters, however, the factor $\sum_{i,j=1}^N e^{\text{i}(\bm{q}_\parallel - \bm{k})\cdot(\bm{r}_j - \bm{r}_i)}$ deviates from the Dirac delta function, and may include in-plane momenta near $\bm{k}$, as clearly illustrated in Fig.~\ref{fig:SJustification_Momentum}(b). In this case, it is reasonable to approximate the resultant momentum distribution of the hybridized state $\zeta(\bm{q}_\parallel,\bm{k})$ with a Gaussian envelope for in-plane momentum space~($\bm{q}_\parallel$).

In this work, we analyze the temporal and spectral evolution of $\alpha_\text{D}(t)$ using two approaches: (i) directly calculating the summation in Eq.~\eqref{SEq:Finite_number} without a Gaussian distribution and (ii) performing the calculation using a Gaussian approximation. In the latter case, namely, by assuming a Gaussian distribution as Eq.~\eqref{SEq:Finite_number} and substituting this expression into Eq.~\eqref{SEq:Dicke_DynamicsGeneral}, we obtain the TDS dynamics as
 \begin{equation}
    \dot{\alpha}_\text{D}(t) \sim -\mathcal{C}'\int_0^t d \tau 
     \int d\tilde{\omega}\int d^2\bm{q_\parallel}~ \zeta(\bm{q}_\parallel,\bm{k})~\mathcal{J}(\tilde{\omega},\bm q_\parallel)  e^{\text{i}(\omega_\text{eg}-\tilde{\omega})(t-\tau)} \alpha_\text{D}(\tau),
    \label{SEq:alpha_dot_First}
 \end{equation}
which is Eq.~\eqref{Eq:Wavefunction_Momentum} of the main text. Equation \eqref{SEq:alpha_dot_First} is still general and applies to a wide class of interactions, including collective emitters and waveguides with arbitrary dissipative emitter-emitter coupling. Nevertheless, Eq.~\eqref{SEq:alpha_dot_First} can be uniquely used for our proposed plasmonic WQED by linearizing the dispersion relation of the plasmonic structure $\omega_\text{spp}(\bm{q}_\parallel)$ around the frequency of interest $\omega_\text{eg}$.

To this aim, we assume that the plasmonic waveguide operates in a single-mode regime and supports a propagative surface-plasmon field whose excitation frequency $\tilde{\omega}_\text{spp}(\bm{q}_\parallel)=\omega_\text{spp}(\bm{q}_\parallel)+i\gamma_\text{spp}$ satisfies the dispersion relation of an extended metal-dielectric interface~(see \S~\ref{Sec:General_SPP} and Eq.~\eqref{Eq:Dispersion_Plasmonic}). Using this dispersion relation and exploiting the Fourier representation of the surface-plasmon field, we find that the Green’s function in complex frequency space contains a single pole at $\tilde{\omega}\text{spp}(\bm{q}\parallel)$, whose residue is given by $\mathcal{A}{ij}(\bm{q}\parallel) := \mathcal{A}(\bm{q}\parallel)$ and exhibits homogeneous linewidth broadening $\gamma\text{spp}(\bm{q}_\parallel)$. We use this framework and corresponding pole contribution to calculate the out-of-plane component of the Green’s function $\bm{G}_\text{zz}\propto\bm{p}_i\cdot\bm{G}\cdot\bm{p}_j$, whose imaginary part, upon using \eqref{SEq:Emitter_Emitter}, determines the emitter-emitter coupling. We then approximate the calculated $\Im[\bm{G}_{zz}]$ in the single-pole regime by a Lorentzian lineshape, and rewrite the emitter-emitter coupling as 
\begin{equation}
    \mathcal{J}(\bm{q}_\parallel,\tilde{\omega})=\frac{\gamma_\text{spp}}{2\pi}\frac{\mathcal{A}(\bm{q}_\parallel)}{(\tilde{\omega}-\omega_\text{spp}(\bm{q}_\parallel))^2+\gamma_\text{spp}^2(\bm{q}_\parallel)}.
    \label{Eq:Semitter_Emitter}
\end{equation}
Following the Fourier-space analysis of the Green’s function at the metal-dielectric interface, $\mathcal{A}(\bm{q}_\parallel)$ for $z,z'>0$ depends on the reflection coefficient. We evaluate this quantity for real-momentum and complex-frequency spaces using the theoretical framework developed in previous investigations~\cite{PhysRevB.79.195414}.

Despite these straightforward quantitative steps, we note that the surface-plasmon excitation frequency is a highly complex function of in-plane momentum due to the inherently nonlinear nature of the dispersion relation. In this work, we linearize this dispersion by assuming that the propagating surface-plasmon mode has a slow group velocity $v_\text{g}$. Furthermore, to simplify the calculations, we consider the resonance coupling $\omega_\text{spp}\approx\omega_\text{eg}$ and, hence, express the plasmonic dispersion as
\begin{equation*}
    \omega_\text{spp}(\bm{q}_\parallel)\approx\omega_\text{eg}+v_\text{g}\cdot\bm{q}_\parallel.
\end{equation*}
We use this linearization, assume $\mathcal{A}(\bm{q}_\parallel)\approx\mathcal{A}(\bm{k})$, and substitute the resultant expression into the emitter-emitter coupling $\mathcal{J}(\bm{q}_\parallel,\tilde{\omega})$~(namely Eq.~\eqref{Eq:Semitter_Emitter}). Subsequently, the integration over complex frequency can be performed by assuming $\omega_\text{spp}(\bm{q_\parallel})\gg\gamma_\text{spp}(\bm{q_\parallel})$ and applying the residue theorem, while the integration over momentum can be computed using $d^2\bm{q}_\parallel = q_\parallel dq_\parallel d\phi$, with $q_\parallel\in[0,\infty)$ and $\phi\in[0,2\pi]$. Here, to perform integration over in-plane momentum, we use the following identity
\begin{equation}
    \int_0^{2\pi} e^{-\text{i}v_\text{g} p_\parallel \cos\phi (t-\tau)} \,d\phi = 2\pi J_0\left[v_\text{g} p_\parallel  (t-\tau)\right],
    \label{Eq:Helper_Bessel}
\end{equation}
for $J_0[\cdot]$, the zeroth-order Bessel function of the first kind. Combining all these approximations and Eq.~\eqref{Eq:Helper_Bessel} we obtain the following closed-form expression for momentum integration,
\begin{equation}
    \int_0^\infty p_\parallel e^{-\alpha p_\parallel^2} J_0(\beta p_\parallel) \,dp_\parallel = \frac{1}{2\alpha}\exp\left(-\frac{\beta^2}{4\alpha}\right).
\end{equation}
We then insert these expressions into \eqref{SEq:alpha_dot_First}, to obtain the transition amplitude of a TDS
\begin{equation}
    \dot{\alpha}_\text{D}(t)=-\Omega_\text{s}\int_{0}^{t}d\tau \alpha_\text{D}(\tau) e^{-v_\text{g}^2(t-\tau)^2/4\mathcal{L}^2} e^{-(\gamma_\text{spp}+ \text{i}\bm{k}. \bm v_\text{g} ) (t-\tau)}, 
    \label{Eq:Dynamic_SM}
\end{equation}
which corresponds to Eq.~\eqref{Eq:Dynamic} of the main text. As is evident, the quantitative steps leading to Eq.~\eqref{Eq:Dynamic_SM} are lengthy and highly mathematical; however, they rely on physically valid technical and methodological approximations. Since these quantitative steps do not contribute novel physical insights into collective-light-collective-matter interaction, we discuss them primarily in the SI. The physical rationale for our derivation of Eq.~\eqref{Eq:Dynamic_SM} is presented in the main text; for self-containment, we summarize these details below.

We note that the steps leading to Eq.~\eqref{Eq:Dynamic_SM} rely on the in-plane translational symmetry, which is strictly valid for an extended metal-dielectric interface~(namely, $\mathcal{L}\to\infty$). In our case, we assume that this condition also holds for our waveguide, whose in-plane dimensions are several times larger than the wavelength of the surface-plasmon field. Furthermore, we assume that the TDS evolution described by Eq.~\eqref{Eq:Dynamic_SM} is also valid for a finite number of quantum emitters~(see Eq.~\eqref{SEq:Finite_number} for mathematical equivalence). This assumption is consistent with previous investigations~\cite{PhysRevResearch.4.023207,masson2022universality} and appears feasible within realistic experimental configurations. In addition to these technical assumptions, our derivation also contains some methodological steps that are unique to our plasmonic waveguides. The first assumption concerns the dispersion relation of the plasmonic waveguide, which we linearize as $\omega_\text{spp}\approx\omega_\text{spp}+\bm v_\text{g}\cdot\bm{q}_\parallel$ around the resonant frequency $\omega_\text{spp}=\omega_\text{eg}$; this assumption ensures that the frequency corresponding to the Dicke-state superradiant is predominantly supported and propagated by the plasmonic waveguide. The second assumption is based on the spectral representation of the plasmonic field in the Fourier domain. Specifically, we calculate the evolution for real momentum and complex frequency, which we exploit in this work. We have justified the validity of this approach and successfully applied it in our quantitative analysis. The dual situation—Fourier optics for complex in-plane momentum and real frequency—has also been developed~\cite{PhysRevB.79.195414}. Following previous investigations, we expect that both approaches yield equivalent results. 

\subsection{\label{Sec:Challenge_Dynamics} Notes on master equation formalism}
In this section, we provide a quantitative description leading to the derivation of the master equation for our proposed plasmonic WQED. Before presenting our analysis, we note that the assumptions and simplifications used in the derivation of Eq.~\eqref{Eq:Dynamic_SM} are also valid for the master equation formalism. In particular, an explicit master equation would require a detailed understanding of the many-emitter interaction of the TDS situated on top of the metallic layer, including the momentum and frequency dependence of the metallic layer in the dynamical evolution, as well as accurate time-step considerations for relaxations within a non-Markovian framework. Hence, deriving a master equation that incorporates many-emitter effects and an accurate description of loss channels is challenging and goes beyond the scope of our current work~(see \S~\ref{Sec:Master_Equation_USC_Regime}). Nevertheless, with the use of approximations and simplifications we made in the main text and here in the SI, developing the master equation framework does not appear to be an ambitious task. Indeed, we expect that such a formalism can be developed for our system when combined with standard assumptions, such as the rotating-wave and Born–Markov approximations. Our aim here is to outline the essential qualitative steps for deriving the master equation when the emitter is prepared in the TDS, and the plasmonic system is initialized in the vacuum state.

We consider the quantum emitters and the plasmonic waveguide as the system and the bath, respectively, and denote their density matrices by $\rho_\text{S}$ and $\rho_\text{B}$. Then evolution of the total density matrix $\rho_\text{SB}(t):=\rho_\text{S}(t)\otimes\bar{\rho}_\text{B}$ for equilibrium system-bath states, where $[H_\text{int}(t),\rho_\text{SB}(0)]=0$, is given by
\begin{equation*}
    \dot{\rho}_\text{SB}(t)=-\frac{1}{\hbar^2}\int_0^t~dt'[H_\text{int}(t),[H_\text{int}(t'),\rho_\text{SB}(t')]].
\end{equation*}
Here $H_\text{int}$ is the interaction Hamiltonian, which, indeed is expressed in a rotated frame, i.e., with respect to $H_\text{e}+H_\text{f}$~(see \S~\ref{Sec:Mathemtics} and Eq.~\eqref{Eq:Hamiltonian_Int}), whose QE and field operators transform as Eq.~\eqref{Eq:Rotated_Sigma} and \eqref{Eq:Rotated_f}, respectively. Following the conventional open-quantum-system framework, we achieve the system evolution by performing a trace over bath states $\dot{\rho}_\text{S}(t)=\text{tr}_\text{B}\{\dot{\rho}_\text{SB}(t)\}$. The dynamics of the TDS can then be expressed as
\begin{equation}
    \dot{\rho}_\text{S}(t)=-\frac{1}{\hbar^2}\int_0^t~dt'\text{tr}_\text{B}\{[H_\text{int}(t),[H_\text{int}(t'),\rho_\text{SB}(t')]]\}.
    \label{Eq:Density_First}
\end{equation}
y assuming the rotating-wave approximation and inserting QEs and field operators, we derive the interaction Hamiltonian, namely, Eq.~ \eqref{Eq:Hamiltonian_Int} as
\begin{equation}
    H_\text{int}(t)=-\text{i}\sqrt{\frac{\hbar}{\pi\varepsilon_0}}\sum_{l=1}^{N}\left(\sigma_+^{(l)}\int_0^{\infty}~d\tilde{\omega}\left(\frac{\tilde{\omega}}{\text{c}}\right)^2\text{e}^{-\text{i}(\tilde{\omega}-\omega_\text{eg})t}
    \int~d^3\bm{r}'\sqrt{\varepsilon_\text{i}(\bm{r}',\tilde{\omega})}\bm{p}_l^*\cdot\bm{G}(\bm{r}_l,\bm{r}',\tilde{\omega})\cdot\bm{f}(\bm{r}',\omega)\right)+\text{c.c.}.
    \label{Eq:S_Hamil_Lindb}
\end{equation}
Then, we obtain the master equation in the same way as in the conventional open-quantum-system framework by inserting Eq.~\eqref{Eq:S_Hamil_Lindb} into Eq.~\eqref{Eq:Density_First}. The quantitative steps required to derive a closed-form master equation would necessitate specifying the statistical properties of the plasmonic field, which is generally challenging~(and indeed an open question) for many-plasmon states.

These steps, however, can be significantly simplified by assuming that the plasmonic system is initialized in its vacuum state. Under this assumption, the following normally ordered bath correlation functions vanish
\begin{align*}
   \text{tr}_\text{B} \{\bm f(\bm{r}',\tilde{\omega})\cdot\bm f(\bm{r}'',\tilde{\omega}')\bm \rho_{B}(0)\} = 0,\quad \text{tr}_\text{B} \{\bm f^\dagger(\bm{r}',\tilde{\omega})\cdot\bm f^\dagger(\bm{r}'',\tilde{\omega}')\bm \rho_{B}(0)\} = 0,\quad \text{tr}_\text{B} \{\bm f^\dagger(\bm{r}',\tilde{\omega})\cdot\bm f(\bm{r}'',\tilde{\omega}')\bm \rho_{B}(0)\} = 0, 
\end{align*}
and the only non-zero bath correlation function reduces to
\begin{equation}
    \text{tr}_\text{B} \{\bm f(\bm{r}',\tilde{\omega})\cdot\bm f^\dagger(\bm{r}'',\tilde{\omega}')\bm \rho_\text{B}(0)\} = \delta(\tilde{\omega}-\tilde{\omega}')\delta(\bm{r}'-\bm{r}'').
\end{equation}
By inserting these results into \eqref{Eq:S_Hamil_Lindb}, introducing the memory kernel
\begin{align}
    C_{ij}(t,t')= \frac{\mu_0}{\pi c^2}
    \sum_{l,l'=1}^N \int d \tilde{\omega}' \int d\bm{r}' \int_0^t dt'  \tilde{\omega}'^4  
      e^{\text{i}(\omega_{eg}-\tilde\omega')(t-t')} 
       \varepsilon_\text{i}(\bm{r}',\tilde{\omega}) \bm{p}_i.\bm{G}(\bm{r}_{l'},\bm{r}';\tilde{\omega}').\bm{G}^*(\bm{r}_{l},\bm{r}';\tilde{\omega}').\bm{p}_j,
 \end{align}
and then making use of Green’s function multiplication identity given in \eqref{Eq:Green_Identity}, we obtain the following master equation
\begin{equation}
    \partial_t\bm\rho_{S}=
    - \sum_{i=1}^N \sum_{j=1}^N  \int_0^t dt'
   ([\bm{\sigma}^+_{i}, \bm{\sigma}^-_{j}  \bm\rho_{S}(t')] C_{ij}(t,t')  - 
    [\bm{\sigma}^-_{i} , \bm\rho_{S}(t')  \bm{\sigma}^+_{j}]  C^*_{ij}(t,t')).
    \label{Eq:Master_Eq}
\end{equation}
We note that although Eq.~\eqref{Eq:Master_Eq} seems quantitatively well-structured, numerical methods for solving this type of integrodifferential equation remain insufficiently explored. In particular, Eq.~\eqref{Eq:Master_Eq} is expressed in matrix form, and the convergence and stability of such a matrix-valued integrodifferential equation must be carefully examined~(see \S~\ref{Sec:Master_Equation_USC_Regime}). Therefore, while Eq.~\eqref{Eq:Master_Eq} provides a rigorous theoretical framework for studying collective light–matter interactions in full detail, its numerical stability and convergence remain uncertain. Consequently, we adopt the Schr\"odinger picture and the scalar integrodifferential equation \eqref{Eq:Dynamic_SM}, for which convergence and stability properties are well established in the mathematical literature.
\end{widetext}

\section{\label{Sec:General_SPP} Fourier analysis of surface-plasmon field}
As discussed in \S~\ref{Derivation_Equations} our theoretical framework for calculating the transition amplitude of the hybridized state $\alpha_\text{D}(t)$~(which is also proportional to the evolution of the HPP for $t>0$) is based on both technical and methodological assumptions. The technical steps enable the use of surface-plasmon Fourier optics to simplify emitter-emitter coupling, specifically by leveraging the in-plane translational symmetry consistent with the momentum-matching condition. The methodological steps further specify our derivation for a nanoplasmonic waveguide by performing a Fourier-space analysis of the surface plasmon in complex frequency space. In this context, we exploit a single-mode plasmonic approximation, whose dispersion~(as a function of in-plane momentum $\bm{q}_\parallel$) can be linearized around the resonant excitation frequency.

To assess the feasibility of our theoretical framework for a nanoplasmonic waveguide and to delineate its physically relevant range of applicability, we introduce experimentally feasible, realistic parameters for the metallic layer. Such parameters can also be specified for QEs. In this work, we assume that the QEs are placed at an optimal distance with respect to $z=0$, and neglect their radiative decays to the far field as well as other non-radiative decays. In what follows, we present only one set of parameters related to the metallic layer; however, we have performed extensive numerical tests over a wide range of metals and identified several sets that can serve as numerical justification~(see \S~\ref{Sec:Physical_Range}). Therefore, the parameter set used here and in the text is not unique.

\begin{figure*}
    \centering
    \includegraphics[width=0.99\linewidth]{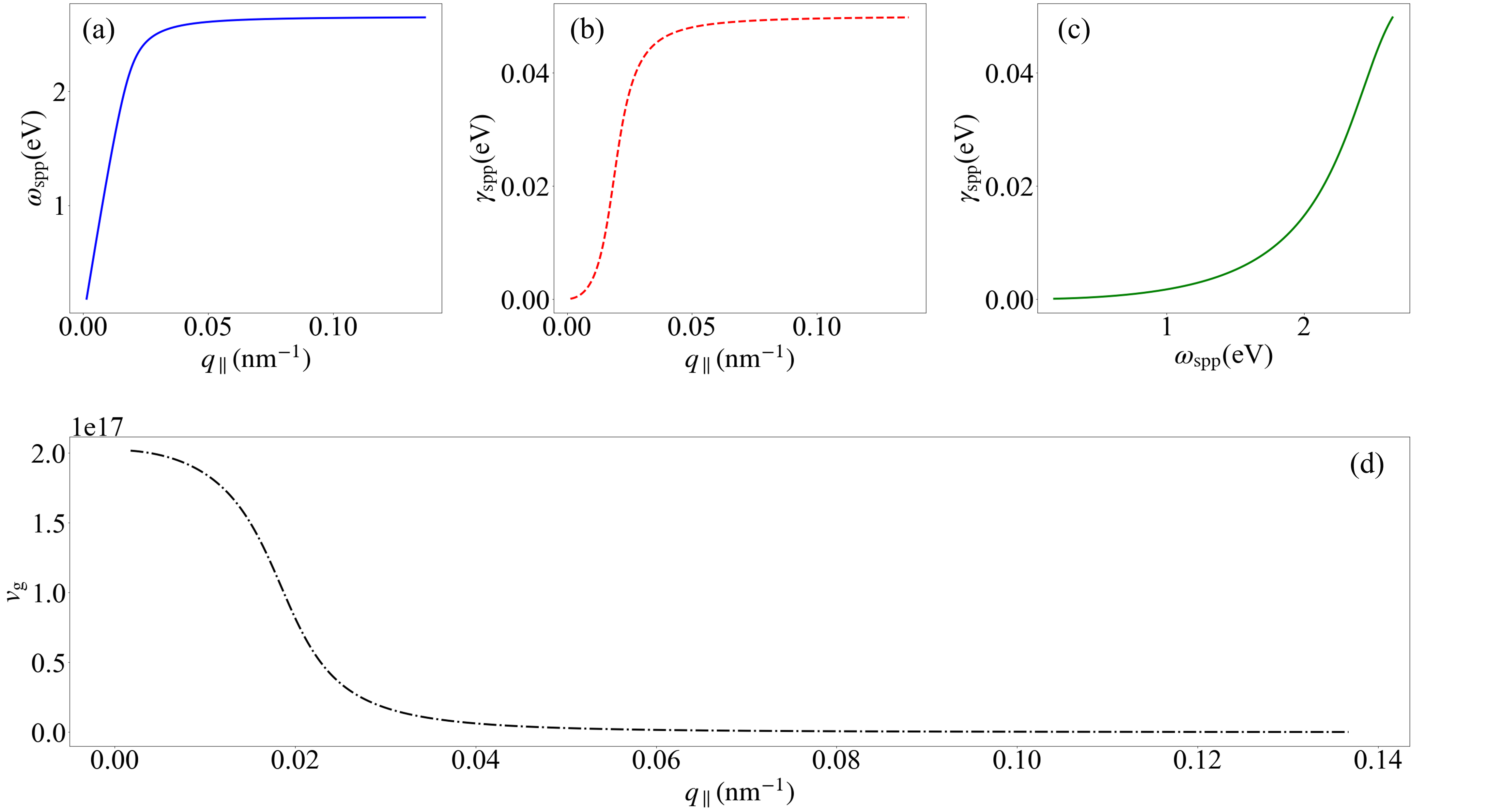}
    \caption{Surface plasmon dispersion $\tilde{\omega}_\text{spp}$ and group velocity for metal-dielectric interface, which is achieved by numerically solving Eq.~\eqref{Eq:Dispersion_Plasmonic}: Panel (a) is the momentum-dependent excitation frequency $\omega_\text{spp}:=\Re[\tilde{\omega}_\text{spp}]$ and panel (b) is its corresponding rate rate $\gamma_\text{spp}:=\Im[\tilde{\omega}_\text{spp}]$. In panel (c) is the complex eigenfrequency curves, and panel (d) is the group velocity of the surface-plasmon polariton field. We plug $\varepsilon_\text{g}=2.2$, $\varepsilon_\infty=9$, $\gamma_\text{pl}=0.1$~eV and $\hbar\omega_\text{pl}=9$~eV into \eqref{Eq:Drude} and solve \eqref{Eq:Dispersion_Plasmonic} for $\tilde{\omega}$. We have used $\tilde{\omega}_0=0.01(1+i)$~eV to solve this nonlinear equation. See the text for more details.}
    \label{fig:SJustification}
\end{figure*}

To illustrate the feasibility of our scheme, we now focus on a single representative set of parameters by choosing gold as the metallic layer. In particular, we consider a thick gold layer that supports a surface-plasmon field characterized by a plasma frequency $\hbar\omega_\text{pl} = 9$~eV, a decay rate $\gamma_\text{pl}=0.1$~eV, and a background permittivity $\varepsilon_\infty=9$. This layer is described by the Drude model
\begin{equation}
    \varepsilon_\text{m}(\bm{r'},\tilde{\omega})=\varepsilon_\infty-\frac{\omega_\text{pl}^2(\bm{r'})}{\tilde{\omega}(\tilde{\omega}+i\gamma_\text{pl}(\bm{r'}))}.
    \label{Eq:Drude}
\end{equation}
In what follows, we neglect the local-field effects and assume Eq.~\eqref{Eq:Drude} is spatially homogeneous for $z<0$; hence $\varepsilon_\text{m}(\bm{r'},\tilde{\omega}):=\varepsilon_\text{m}(\tilde{\omega})$. We then assume that a dielectric layer with permittivity $\varepsilon_\text{g}=2.2$ is placed on top of this metallic structure. This choice of parameters is exemplary, as outlined above; indeed, we have examined various values of permittivity within the range $\varepsilon_\text{g}\in[1.45,3.45]$ and observed similar surface-plasmon dispersion relation~(as in Fig.~\ref{fig:SJustification}(a) and \ref{fig:SJustification}(b)) and resonant evolution in the complex frequency plane~(shown in Fig.~\ref{fig:SJustification}(c)).

Following the conventional electromagnetic boundary condition for a single metal-dielectric interface~\cite{PhysRev.182.539}, and by defining the $z$-component of the momentum wavevector as  
\begin{equation*}
    k_\jmath = \sqrt{\left|\bm{q}_\parallel\right|^2-\left(\frac{\tilde{\omega}}{\text{c}}\right)^2\varepsilon_\jmath},
\end{equation*}
for $\jmath\in\{\text{m},\text{g}\}$, we obtain the surface-plasmon dispersion relation by numerically solving the characteristic equation
\begin{equation}
    \frac{k_\text{m}(\tilde{\omega},\bm{q}_\parallel)}{\varepsilon_\text{m}(\tilde{\omega})}+\frac{k_\text{g}(\tilde{\omega},\bm{q}_\parallel)}{\varepsilon_\text{g}}=0,
    \label{Eq:Dispersion_Plasmonic}
\end{equation}
which we solved for real in-plane momentum $\bm{q}\parallel$ and complex frequency $\tilde{\omega}$. We represent the resultant numerical solution of the surface-plasmon excitation $\omega_\text{spp}$ obtained from Eq.~\eqref{Eq:Dispersion_Plasmonic} in Fig.~\ref{fig:SJustification}(a) and display the surface-plasmon decay rate $\gamma_\text{spp}$ in Fig.~\ref {fig:SJustification}(b), both plotted as functions of  $\bm{q}_\parallel$. Furthermore, we illustrate the evolution of the surface-plasmon excitation in the complex-frequency plane $(\omega_\text{spp},\gamma_\text{spp})$~(shown in Fig.~\ref{fig:SJustification}(c)). We also identify the excitation of the slow-mode surface-plasmon field, as shown in Fig.~\ref{fig:SJustification}(d).

In particular, we obtain $$\omega_\text{spp}(\bm{q_\parallel\to\infty})=\frac{\omega_\text{pl}}{\sqrt{\varepsilon_\infty+\varepsilon_\text{g}}}\approx2.55~\text{eV}$$ as the asymptotic frequency (see the $q_\parallel\to\infty$ limit in Fig.~\ref{fig:SJustification}(a)). For an exemplary quantum emitter with transition frequency $\omega_\text{eg}=1.5$~eV, we find a finite in-plane momentum $q_\parallel=0.012~\text{nm}^{-1}$ and a relatively low plasmonic field decay $\gamma_\text{spp}\approx0.005$~eV. In this scenario, we achieve a slow-plasmon field with group velocity $v_\text{g}\approx1.768\times10^{17}~\text{nm}\cdot\text{s}^{-1}$ (namely, $v_\text{g}\approx0.59\text{c}$) propagating along the interaction interface. The frequency evolution and group velocity of this propagating surface-plasmon field satisfy the collective-matter criteria presented in the main text and are suitable for investigating the temporal and spectral evolution of the hybridized state.

We then introduce the $p$-polarized Fresnel reflection coefficient~\cite{PhysRevB.79.195414} as
\begin{equation*}
    r_\text{p}=\frac{k_\text{g}\varepsilon_\text{m}-k_\text{m}\varepsilon_\text{g}}{k_\text{g}\varepsilon_\text{m}+k_\text{m}\varepsilon_\text{g}}.
\end{equation*}
By inserting numerical values for the in-plane momentum and complex frequencies as outlined above, we obtain $r_\text{p}\approx0.882+i0.47$. Finally, by utilizing $p$-polarized unit vectors, 
\begin{equation}
    \bm{p}_1^\pm = \frac{|\bm{q}_\parallel|\bm{z}\mp k_\text{g}\bm{q}_\parallel/|\bm{q_\parallel}|}{k_0\sqrt{\varepsilon_\text{d}}},
\end{equation}
expanding the Green’s function at the interface as
\begin{equation}
    \bm{G}(z_i,z_j;\bm{q}_\parallel,\tilde{\omega})=\frac{\text{i}}{2k_\text{g}}[\bm{p}_1^- r_\text{p}\bm{p}_1^{+}]\exp\{\text{i}k_\text{g}(z_i-z_j)\}
\end{equation}
and assuming $z_i=z_j$, we obtain the out-of-plane component of the Green’s tensor as $$|\bm{G}_{ij}(z_i,z_j;\bm{q}_\parallel,\tilde{\omega})|\approx 1.024\times10^{10}~\text{nm}.$$ We note that in the range of dielectric permittivity $\varepsilon_\text{g}\in[1.45,3.45]$, the out-of-plane Green’s tensor component varies from $4.018\times10^9$ to $2.602\times10^{10}$. However, its numerical values do not significantly affect the temporal dynamics of the hybridized state. Therefore, we fix this parameter at $|\bm{G}_{zz}(z_i,z_j;\bm{q}_\parallel,\tilde{\omega})|\approx 1\times10^{10}~\text{nm}$ and treat the number of quantum emitters as a tunable parameter.

\section{\label{Sec:Different_Lattice} Emitter Configuration}
\subsection{Numerical analyses for finite number of quantum emitters}
As outlined in the main text, we do not consider TDS excitation as a quantum-state preparation problem. Such state preparation would require a careful microscopic study, including the many-body evolution of both the quantum emitters~(which may exhibit non-equilibrium dynamics during the preparation steps) and the metallic layer~(treated as a many-electron system). Without going into the details, we assume that the TDS is prepared by an external coherent field~(with momentum $\bm{k}$) before the interaction between the TDS and the surface-plasmon field is initialized~(e.g., at $t_0\in(-\infty,0]$). Despite the complexity in TDS excitation, the derivation of Eq.~\eqref{Eq:Dynamic_SM}, however, depends on the spatial distribution of the QEs and whether they satisfy the phase-matching (momentum-matching) condition. Disregarding many-body corrections to the quantum-state preparation, the effect of emitter distribution on the temporal evolution of $\alpha_\text{D}(t)$ can be analytically investigated using the effective phase relation between all emitters, characterized by $\sum_{ij=1}^N\exp\{\text{i}(\bm{q}_\parallel-\bm{k})\cdot (\bm{r}_i-\bm{r}_j)\}$.

In this section, we aim to explore how the spatial arrangement and the number of emitters affect the transition amplitude of the hybridized state $\alpha_\text{D}(t)$. The differences in temporal and spectral evolutions related to $N$ arise from the fact that all emitters are mutually coupled via the surface-plasmon Green’s function. In this all-to-all interaction context,
\begin{equation*}
    \sum_{ij=1}^N\exp\{\text{i}(\bm{q}_\parallel-\bm{k})\cdot (\bm{r}_i-\bm{r}_j)\},
\end{equation*}
shall be used instead of an approximate Gaussian distribution function. To investigate the effect of emitters' configuration on $\alpha_\text{D}(t)$, we use Eq.~\eqref{SEq:Dicke_DynamicsGeneral}, and consider many-particle arrangements in different lattice geometries.

\begin{figure*}
    \centering
    \includegraphics[width=0.8\linewidth]{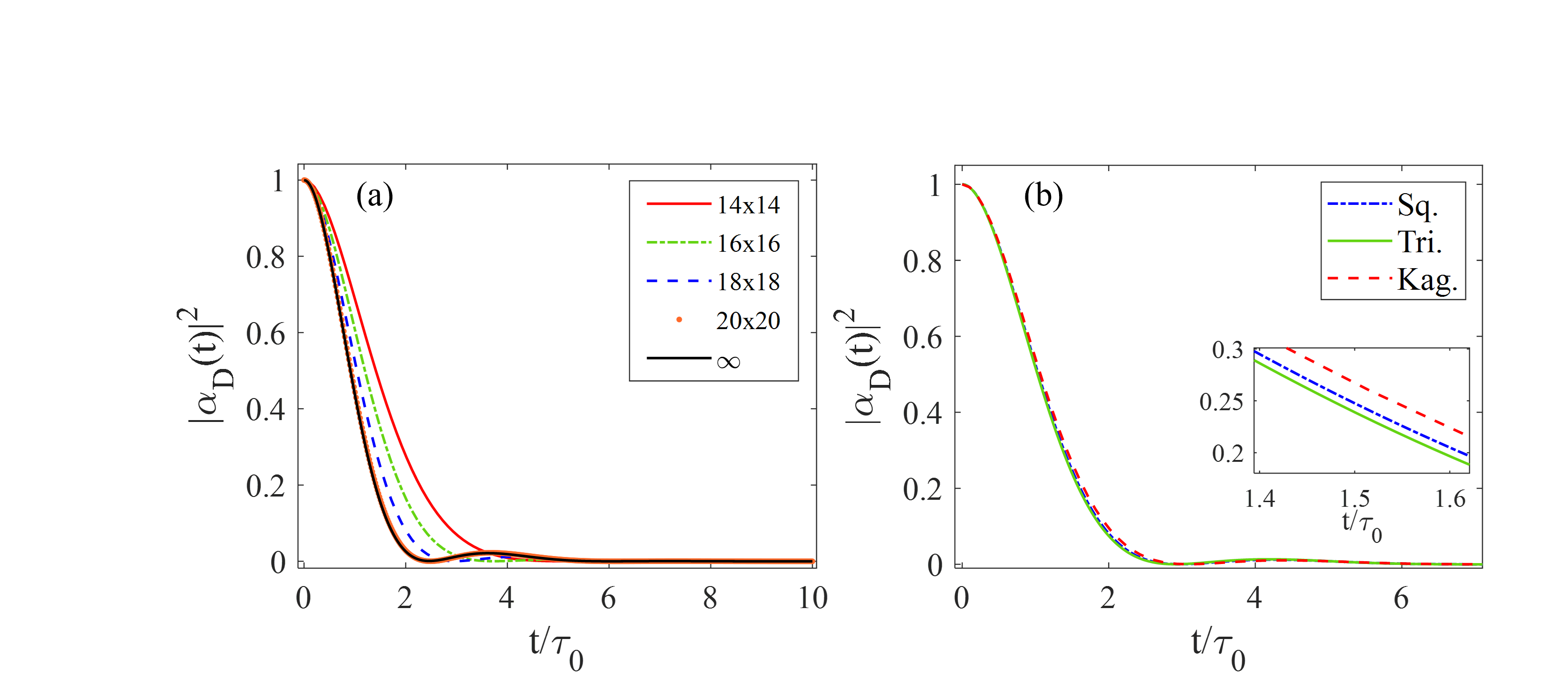}
    \caption{Non-Markovian temporal evolution of $\alpha_\text{D}(t)$ as described by Eq.~\eqref{Eq:Dynamic} for different numbers of emitters. (a) Comparison of TDS dynamics for various lattice sizes with the Gaussian distribution approximation given in Eq.~\eqref{Eq:Dynamic_SM}. (b) Effect of lattice geometry—square, triangular, and Kagome—on the TDS for the same number of emitters ($N=18\times18$). In all cases, $\Omega_\text{s}/\gamma_\text{spp} = 1$ is assumed.}
\label{Fig:Temporal_Dynamicslattice}
\end{figure*}

In this analysis, we use $\bm{k}\cdot\bm{v}_\text{g} \sim 0.07$ as an exemplary value for our numerical calculations. For a relatively large number of emitters ($N=20\times20$) and $\Omega\text{s}/\gamma_\text{spp} =1$, we observe good convergence between the transition amplitude dynamics calculated using Eq.~\eqref{SEq:Dicke_DynamicsGeneral} and that obtained from the Gaussian distribution approximation in \eqref{SEq:alpha_dot_First}, as shown in Fig.~\ref{Fig:Temporal_Dynamicslattice}(a). However, for the same lattice configuration with a smaller number of emitters, $N=14\times14$ (red solid line in Fig.~\ref{Fig:Temporal_Dynamicslattice}(a)), the TDS dynamics shifts from critical oscillation to pure decay. To gain deeper insight, we then consider a relatively large number of QEs, $N=18\times18$, arranged in different lattice configurations, specifically square, triangular, and kagome lattices (Fig.~\ref{Fig:Temporal_Dynamicslattice}(b)). We re-solve the integration in Eq.~\eqref{SEq:Dicke_DynamicsGeneral} and obtain the dynamical evolution of $\alpha_\text{D}(t)$. Our numerical analysis confirms that the lattice geometry has no noticeable impact on the HPP’s dynamical evolution. The minor differences are mostly related to the negligible slope of the HPP dynamics, as shown in the inset of Fig.~\textcolor{blue}{S2}(b). Across all lattice geometries, we confirm that a decrease in the number of emitters corresponds to HPP dynamics with a reduced oscillation frequency.

\subsection{Range of validity for number of quantum emitters}
Let us now discuss the domain of validity for the number of QEs. Our choice of the number of QEs, $N$, in the analysis outlined above is motivated by the well-known behavior of Dicke-state dynamics in traditional QED setups, where the TDS emerges prominently for subwavelength emitters satisfying $N\gg1$. In our system, $N$ is bounded from above, since an infinite number of emitters cannot be accommodated within the limited subwavelength area~[see \S~\ref{Sec:Large_N}], and we also expect a lower cut-off for $N$. Regarding the upper limit, our theory applies to a subwavelength, point-like QE distribution, where many-body effects—such as strong emitter correlations and shape effects of the QEs—can be neglected. To characterize the large-$N$ limit, we performed extensive numerical analyses for various values of $N$ and found that a $20\times20$ lattice provides physically meaningful and numerically stable results, with absolute and relative tolerances as low as $10^{-6}$ when converting and solving the integrodifferential equation as an ordinary differential equation. While lower $N$ may yield faster convergence, as shown in Fig.~\ref{Fig:Temporal_Dynamicslattice}(a), the oscillations may deviate from the expected HPP dynamics. Therefore, the number of emitters should be as large as possible to satisfy momentum conservation, but it cannot be taken to the $N\to\infty$ limit due to size constraints and strong emitter-emitter correlations.

Despite these qualitative analyses in the large-$N$ limit, attempting to determine a precise cut-off number $N_\text{cut-off}$ using the same arguments may lead to inaccuracies. In principle, $N_\text{cut-off}$ can be estimated in two steps. First, one may start from a sufficiently large number of emitters $N$ that reproduces the HPP dynamics and gradually reduce $N$ until noticeable changes appear in the dynamics. Second, one could begin with a very small $N$, include all emitter–emitter interactions, and perform the summation over all lattice sites exactly (without using the Gaussian approximation), then gradually increase $N$. The value at which these two approaches converge would define the exact cut-off number $N_\text{cut-off}$. In our work, we follow the first approach and do not implement the second, since we neglect many-body emitter–emitter interactions. Consequently, the cut-off number reported as $14\times14$ is determined numerically and should not be considered rigorously proven. While a fully rigorous cut-off would be ideal, our formalism provides only an approximate estimate.

To conclude this section, we note that for lower $N$, interference effects may become dominant. In this regime, the Fourier-space analysis outlined above may no longer be valid, as interference effects cannot be neglected. Instead, one must derive the dynamical equation of motion for the transition amplitude of each emitter and solve the resulting system of coupled integrodifferential equations. In this context, Ref.~\cite{10.1063/5.0217702} investigates a one-dimensional plasmonic configuration with a very small number of emitters ($N=3$) and demonstrates that the dynamics of each emitter depend on the others, with the total system dynamics obtained by solving the coupled equations. Extending this approach to a two-dimensional configuration and comparing the results with Eq.~\eqref{SEq:Dicke_DynamicsGeneral} could, in principle, allow one to characterize $N_\text{cut-off}$. However, such a study would require a careful examination of the resulting solutions and a detailed comparison between the two approaches—an analysis beyond the scope of the present work.

\subsection{\label{Sec:Large_N}Asymptotic large-\emph{N} limit}
We conclude our analysis of finite-size effects in the QE array by examining the validity of the all-to-all interaction assumption in the asymptotic large-\emph{N} limit~[namely, $N\to\infty$]. In this limit, we provide both qualitative and quantitative analysis of the relation between the scaling of the effective extent $\mathcal{L}$ and the number of QEs, $N$. 

Qualitatively, our derivation of Eq.~\eqref{Eq:Dynamic} is based on the ideal TDS, whose dynamics can be described via the effective frequency $\Omega_\text{s}\sim N\mathcal{J}(\tilde{\omega}_\text{spp},\bm{k})/(2\mathcal{L})^2$, and is valid in the limits $N\gg1$ and $d\to0$. Apart from the memory kernel in Eq.~\eqref{Eq:Dynamic}, the light–matter coupling constant in a conventional WQED system, $\Omega_{\text{s,WQED}}$, is proportional to the dissipative emitter–emitter coupling $\mathcal{J}(\omega_\text{WQED})$~\cite{PhysRevA.95.033818}; namely, $\Omega_{\text{s,WQED}}\propto\mathcal{J}(\omega_\text{WQED})$, where $\omega_\text{WQED}$ denotes the excitation frequency of the waveguide mode. We anticipate that a similar argument applies to the plasmonic WQED case. Therefore, within the expression for $\Omega_\text{s}$, $N$ and $\mathcal{L}^2$ must scale comparably, i.e., $N/(2\mathcal{L})^2\sim\mathcal{O}(1)$, which implies $N\propto\mathcal{L}^2$.

The above intuitive argument, based on conventional WQED, can also be verified quantitatively for plasmonic WQED in the asymptotic large-\emph{N} limit of a square lattice of QEs. We start from Eq.~\eqref{SEq:Finite_number} and define $\delta\bm{k}:=\bm{q}_\parallel-\bm k$~(see Fig.~\ref{fig:SJustification_Momentum}(b)). By rewriting $\sum_{ij=1}^N\exp\{\text{i}(\bm{q}_\parallel-\bm{k})\cdot (\bm{r}_i-\bm{r}_j)\}$ as 
\begin{equation*}
    \sum_{ij=1}^N\text{e}^{\text{i}\delta\bm{k}\cdot(\bm{r}_i-\bm{r}_j)} = \sum_{i=1}^N\text{e}^{\text{i}\delta\bm{k}\cdot\bm{r}_i}\sum_{j=1}^N\text{e}^{-\text{i}\delta\bm{k}\cdot\bm{r}_j},
\end{equation*}
and substitute the subsequent equation into Eq.~\eqref{SEq:Finite_number}, we obtain
\begin{equation}
    \zeta(\bm{q}_\parallel,\bm{k})=\left|\sum_{i=1}^{N}\exp\{\text{i} \delta\bm{k}\cdot\bm{r}_i\}\right|^2.
    \label{Eq:Localization_Middle}
\end{equation}
Next we define $\bm{x}$ ($\bm{y}$) as the unit vector along the $x$~($y$) direction and denote the number of emitters along each axis as $N_x$~($N_y$), to characterize quantum emitter positions by $\bm{r}_l=d(n\bm{x}+m\bm{y})$, with $l\in\{i,j\}$. Substituting these definitions into Eq.~\eqref{Eq:Localization_Middle}, we achieve
\begin{equation}
    \zeta(\bm{q}_\parallel,\bm{k})= \left|\sum_{n=1}^{N_x} \text{e}^{\text{i}\delta \bm{k}_x n d } \sum_{m=1}^{N_y} \text{e}^{\text{i}\delta \bm{k}_y m d } \right|^2,
\end{equation}
which, upon using the identity
\begin{equation*}
    \sum_{l=0}^{N-1} e^{\text{i}l\theta } = e^{i \theta(N-1)/2} \frac{\sin (N\theta/2 ) }{\sin (\theta/2)},
\end{equation*}
yields the following expression for localization
\begin{equation}
    \zeta(\bm{q}_\parallel,\bm{k})=\left[\frac{\sin (N_x\delta\bm{k}_x d/2 ) }{\sin (\delta\bm{k}_x d/2)}\right]^2\left[\frac{\sin (N_y\delta\bm{k}_y d/2 ) }{\sin (\delta\bm{k}_y d/2)}\right]^2.
    \label{Eq:Advanced_Localization}
\end{equation}
We now assume $\delta\bm{k}_l \ll N\delta\bm{k}_l \ll 1$, and use $\sin(\delta\bm{k}_l d/2)\approx\delta\bm{k}_l d/2$ and
\begin{equation*}
    \sin (N_l\delta\bm{k}_l d/2)\approx N_l\delta\bm{k}_l d/2-(N_l\delta\bm{k}_l d/2)^3/3!
\end{equation*}
into Eq.~\eqref{Eq:Advanced_Localization}. On the other hand, we also use the following approximation 
\begin{equation*}
    \zeta(\bm{q}_\parallel,\bm{k})=N^2\exp\{-\mathcal{L}^2(\bm{q}_\parallel-\bm{k})^2\}\approx N^2(1-\mathcal{L}^2\delta\bm{k}^2).
\end{equation*}
Equating these Taylor expansions, keeping terms up to $\delta\bm{k}_l^2$, and using $N=N_xN_y$, we obtain the final result
\begin{equation}
    N\sim 3(\mathcal{L}/d)^2,
\end{equation}
which confirms the earlier intuitive elucidation based on conventional waveguide QED.

\section{Discussion and Outlook}
In this section, we outline the outlook and possible future directions related to the waveguide-mediated nanoscopic collective light–matter interaction. We first delineate the physical range of applicability of our proposed concept and discuss the similarities and distinctions between our collective light–matter effects in plasmonic WQED and analogous platforms such as cavity QED. We then provide a guideline for the experimental feasibility of our scheme, and finally, we discuss prospective research avenues and the broader panorama of nanoscopic collective light–matter interaction.

\subsection{\label{Sec:Physical_Range} Physical range of applicability}
In this work, we investigate the interaction between a TDS and a surface-plasmon field, aiming to explore collective light–matter effects at subwavelength scales. Our key parameter of interest is $\alpha_\text{D}(t)$. Owing to the hybridized nature of the interaction, $\alpha_\text{D}(t)$, in principle, depends on both the surface-plasmon parameters $(\gamma_\text{spp}, \omega_\text{spp})$ and the TDS parameters $(N, \gamma_\text{D}, \omega_\text{eg}, \bm{k})$. However, this dependency is complicated due to the non-Markovian evolution of the HPP. In this work, we perform a mapping that converts the integrodifferential equation into a damped harmonic oscillator and show that the temporal and spectral evolution of $\alpha_\text{D}(t)$ can be effectively characterized by an oscillation frequency $\omega_\text{s}$ and a long-time decay rate $\Gamma_\text{s}$. We then introduce the collective complex frequency $\tilde{\omega}_\text{s} = \omega_\text{s} + \text{i}\Gamma_\text{s}$ to describe the HPP evolution.

\begin{figure}
    \centering
    \includegraphics[width=0.65\linewidth]{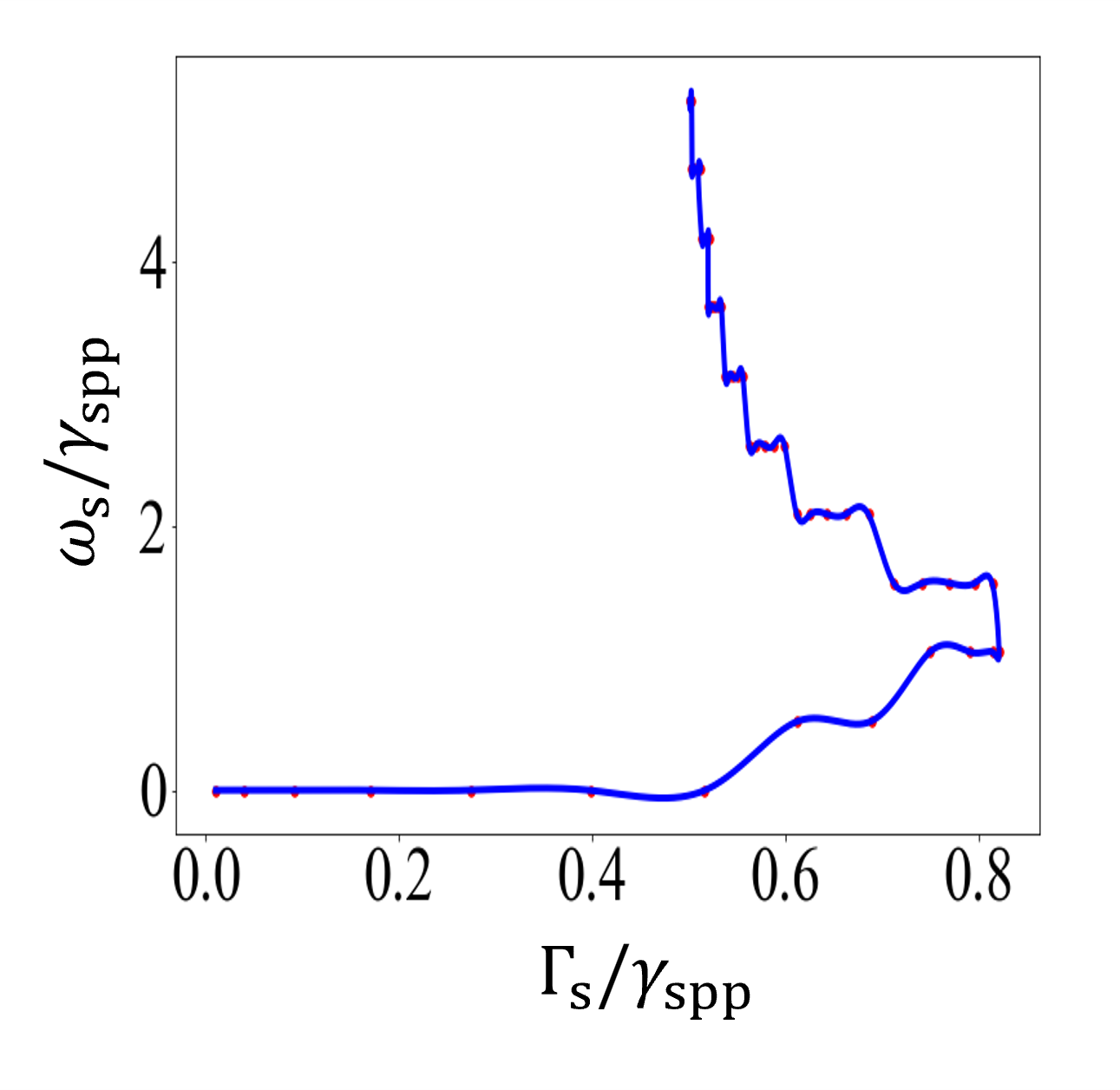}
    \caption{Evolution of collective oscillations in the complex frequency plane $(\omega_\text{s},\Gamma_\text{s})$. The curve defines three regions: pure decay ($\Gamma_\text{s}/\gamma_\text{spp}<0.5$) with no oscillation, an intermediate range ($\Gamma_\text{s}/\gamma_\text{spp}\in[0.5,1]$) corresponding to critical oscillation where long-time decay increases with collective oscillation, and $\omega_\text{s}/\gamma_\text{spp}>1$, where long-time decay decreases and asymptotically approaches $\Gamma_\text{s}=0.5\gamma_\text{spp}$. Numerical values are the same as those used in Fig.~\ref{fig:Geometry_Waveguide} of the main text.}
\label{Fig:Strong_Weak_Coupling}
\end{figure}

This harmonic-oscillator interpretation of our interaction provides a clearer understanding of the different coupling regimes, even in the presence of non-Markovian effects. In particular, pure and oscillatory decays—corresponding to weak, weak-to-strong, and strong coupling regimes—can be distinguished by comparing the collective HPP frequency with its long-time decay, a situation also observed in conventional cavity-QED configurations~\cite{PhysRevB.87.115419}. Specifically for our system, we introduce the parameter $\omega_\text{s}\approx\Omega_\text{s}$~(see the values of $\omega_\text{s}$ for different $$\Omega_\text{s}\in\{1,2,3,4\}$$ for justification) and the long-time decay rate $\Gamma_\text{s}$ for the dynamics of $\alpha_\text{D}(t)$, which together can be used to control and tune the coupling strength. Accordingly, weak, strong, and intermediate~(weak-to-strong) coupling regimes are accessible by adjusting the effective frequency relative to the long-time decay. In analogy with the cavity-QED framework, oscillatory decay dynamics for $\alpha_\text{D}(t)$ emerge when $\omega_\text{s}(\Omega_\text{s})>\Gamma_\text{s}(\Omega_\text{s})$, whereas a purely decaying response occurs when $\omega_\text{s}(\Omega_\text{s})<\Gamma_\text{s}(\Omega_\text{s})$, as illustrated in Fig.~\ref{fig:Spectral_Dynamcis}(a) and Fig.~\ref{Fig:Strong_Weak_Coupling}. We note that in the oscillatory decay regime, the system exchanges energy between the eigenstates $\ket{\Psi_\text{D}(N|\bm{k})}$ and $\ket{\psi_\text{G}}$~(see Eq.~\eqref{Eq:Wavefunction}) of the main text for the definition of these states), arising from vacuum effects, which is a hallmark of strong coupling.

Aforementioned discussion justifies, quite generally, that the collective light-matter interaction in our plasmonic WQED is fundamentally a time-scaling problem, which can be investigated either using Eq.~\eqref{Eq:Dynamic_SM} or the damped harmonic-oscillator representation~(Eq.~(\textcolor{blue}{8}) in the main text). In the original frame, characterized by integrodifferential equation~(Eq.~\eqref{Eq:Dynamic_SM}), we define the memory time as $\tau_\text{mem}=\sqrt{2}\mathcal{L}/v_g$ and the coupling time as $\tau_\text{cou}:=1/\Omega_\text{s}$. The HPP’s evolution can then be identified by the interplay between $\tau_\text{mem}$, $\tau_\text{cou}$, and $\tau_0:=1/\gamma_\text{spp}$. Assuming slow surface-plasmon field excitation, $v_g \approx0.1\mathcal{L}\gamma_\text{spp}$, we obtain $\tau_\text{mem}=10\sqrt{2}\tau_0\gg\tau_0$; thereby, the HPP dynamics depend on the competition between $\tau_\text{cou}$ and $\tau_0$. In this basis, we obtain weak coupling for $\tau_\text{cou}>\tau_0$ and we observe strong coupling for $\tau_\text{cou}<\tau_0$. Using the damped-harmonic representation~(where memory time is long enough to establish energy exchange among the surface-plasmon field and the TDS), weak and strong coupling can be identified similarly to cavity QED assumptions, by defining $\tau_0'=\Gamma_\text{s}$ and $\tau_\text{cou}'=1/\omega_\text{s}$. In particular, we obtain weak coupling for $\tau_\text{cou}'>2\tau_0'$, strong coupling can be achieved for $\tau_\text{cou}'<2\tau_0'$, and we observe weak-to-strong coupling for $\tau_\text{cou}'=2\tau_0'$.

\subsection{Outlook}
We conclude this work by elucidating the detailed possible outlooks and future directions. To this end, in \S~\ref{Sec:Dissimilar_Cavity_QED} we discuss the differences between our work and the conventional cavity-QED framework, highlighting the challenges and opportunities for the experimental realization of our scheme. Next, in \S~\ref{Sec:Master_Equation_USC_Regime}, we examine the potential transition to the ultra-strong coupling regime and emphasize the importance of deriving an appropriate master equation for our system.

\subsubsection{\label{Sec:Dissimilar_Cavity_QED} Dissimilarities to cavity QED}
In \S~\ref{Sec:Physical_Range}, we discuss the similarities between our system and conventional cavity QED, which arise from the fact that the interaction between collective light and collective matter can be investigated as a time-scaling problem, similar to those studied in cavity QED. Despite these similarities, plasmonic WQED preserves in-plane translational symmetry, allowing eigenfrequencies to be identified via a dispersion relation in terms of a continuous wavevector $\bm{k}$, whereas this symmetry is broken in plasmonic cavities, which mostly support discrete eigenfrequencies. The memory kernel~($K(t-\tau)$ in Eq.~\eqref{Eq:Dynamic} of the main text) emerges in the plasmonic WQED, and thus can induce anomalies in long-time decay, $\Gamma_\text{s}$, or in the HPP’s collective frequencies, $\omega_\text{s}$, which are quite dissimilar to those in conventional cavity QED.

In particular, by investigating the HPP’s collective frequency~($\omega_\text{s}$) for various values of the long-time decay, we observe that the effective-frequency increment enhances the long-time decay for $\Omega_\text{s}/\gamma_\text{spp}\approx1$, after which it decreases and asymptotically approaches the surface-plasmon decay at sufficiently large $\Omega_\text{s}$, as shown in Fig.~\ref{fig:Spectral_Dynamcis}. This anomalous behavior emerges in the frequency domain, characterized by the $(\omega_\text{s},\Gamma_\text{s})$ space, where we numerically identify three distinct long-time decay regimes: (i) $\Gamma_{\text{s}}<0.5\gamma_\text{spp}$ corresponds to weak coupling and pure decay; (ii) $0.5\gamma_\text{spp}<\Gamma_{\text{s}}<\gamma_\text{spp}$ corresponding to intermediate coupling, where an increase in the collective frequency yields long-time decay enhancement; and, surprisingly, (iii) $\Gamma_{\text{s}}>\gamma_\text{spp}$ in which an increment in the collective frequency leads to a \emph{reduction} in the long-time loss $\Gamma_\text{s}$. Unique to our plasmonic system, for high collective frequencies, $\omega_\text{s}\geq4\gamma_\text{spp}$, the long-time decay approaches the surface-plasmon decay and becomes independent of the system parameter $\Omega_\text{s}$, as shown in Fig.~\ref{Fig:Strong_Weak_Coupling}. We leave a detailed investigation of the physical origin and a quantitative analysis of the system behavior at high $\omega_\text{s}$ for future work.

Next, as discussed in the main text, we focus on a flat metallic layer and on QEs prepared in the TDS as the primary representatives of collective light and collective matter, respectively. We introduce the collective light–collective matter interaction in its simplest form and neglect effects such as plasmon-field inhomogeneity, deviations of the QEs from the lattice structure, and additional loss mechanisms, including nonradiative plasmonic decay ($\gamma_\text{NR}$) and intrinsic QE decay ($\gamma_\text{int}$). We note that, for a flat metallic layer, the nonradiative decay rate generally depends on the distance between the quantum emitters and the metallic interface~\cite{Chance1978}, and it dominates when the emitters are located within approximately 10~nm of the surface. Nevertheless, reducing the emitter–interface separation increases the Purcell factor and can enhance the TDS decay rate, which in turn suppresses the collective light excitation.

Efficient collective light–matter hybridization therefore requires optimization of the number of emitters $N$, the Purcell factor, emitter–emitter interactions, and nonradiative decay, all within a non-Markovian framework. This optimization does not alter the fundamental concept of collective–collective hybridization; however, it can modulate the strength of the collective light–matter interaction, either enhancing or suppressing it. A detailed analysis requires knowledge of the emitter–emitter interactions, the specific energy-level structure of the emitters, and the spatial inhomogeneity of the plasmonic field. These effects lie beyond the standard cavity-QED formalism, and incorporating them into our framework to address the resulting optimization problem would be desirable; however, doing so would require extensive numerical analysis beyond the scope of the present study. Such investigations could be pursued in future studies of open quantum systems.

\subsubsection{\label{Sec:Master_Equation_USC_Regime} Master equation formalism: challenges and opportunities}
In \S~\ref{Sec:Challenge_Dynamics}, we derive a master equation capable of describing the collective light–matter interaction. Our equation follows the same quantitative framework used to describe the non-Markovian dynamics of open quantum systems~\cite{RevModPhys.89.015001}. Specifically, our non-Markovian master equation [Eq.~\eqref{Eq:Master_Eq}] shares the same mathematical structure as the master equations presented in \S~IV B of Ref.~\cite{RevModPhys.89.015001}. By addressing issues related to convergence and dynamical stability, one can use our master equation to investigate the coherence, correlations, and quantum properties of the system. In this work, we restrict ourselves to the vacuum state of the plasmonic field. Extending beyond this limit would allow exploration of the full spectral and temporal properties of Dicke ladders~\cite{PhysRevA.96.023863,PhysRevA.98.063815}. In particular, various dephasing and pumping processes can be incorporated as Lindblad terms to assess the feasibility of light–matter interaction under more realistic experimental conditions. Exploring the quantum phenomena underlying collective light–matter interactions using our master equation formalism thus represents a promising direction for future research.

As a final note, we demonstrate in this work that different coupling regimes—weak, intermediate (weak-to-strong), and strong—can be achieved in our system by tuning the effective frequency $\Omega_\text{s}$. The coupling strength can also be enhanced to reach the ultra-strong coupling regime~\cite{PhysRevResearch.5.033002} when the well-defined system parameter $\Omega_\text{s}$ satisfies $\Omega_\text{s}/\omega_\text{eg}\simeq0.1$~\cite{frisk2019ultrastrong}. Although this condition is challenging to satisfy in plasmonic WQED due to the intrinsically lossy nature of surface and gap plasmons, it may be approached by reducing the group velocity of the surface plasmon as much as possible and by increasing the number of QEs to enhance the emitter–plasmon coupling strength. In addition, the interaction may need to be triggered by a virtual photon~\cite{de2017virtual}, such as a vibrational excitation~\cite{PhysRevLett.134.063602}, which can modify the ground-state plasmon statistics and enable multi-plasmon energy exchange in the context of collective light–matter interaction. A detailed investigation of the feasibility of ultra-strong coupling in our proposed scheme is reserved for future work.

\bibliography{ref}

\end{document}